\documentclass[%
 aip,
 amsmath,amssymb,
 reprint,%
]{revtex4-1}

\usepackage{graphicx}
\usepackage{dcolumn}
\usepackage{bm}

\usepackage[utf8]{inputenc}
\usepackage[T1]{fontenc}
\usepackage{mathptmx}
\usepackage{etoolbox}
\usepackage{empheq}
\usepackage{amsfonts}
\usepackage{amssymb}
\usepackage{mathtools}
\usepackage{tensor}
\usepackage{numprint}
\usepackage{siunitx}
\usepackage{subfigure}
\usepackage{enumitem}
\usepackage{stackengine}
\usepackage{natbib}
\usepackage[version=4]{mhchem} 
\usepackage[dvipsnames]{xcolor}
\usepackage{amsmath}

\newcommand{\oxchem}{Physical and Theoretical Chemistry Laboratory, Department of Chemistry, University of Oxford, Oxford OX1 3QZ, United Kingdom.}

\newcommand{\leeds}{School of Chemistry, University of Leeds, Leeds LS2 9JT, United Kingdom}
\newcommand{\bristol}{School of Mathematics, University of Bristol, Fry Building, Woodland Road, Bristol, BS8 1UG, United Kingdom}

\makeatletter
\def\@email#1#2{%
 \endgroup
 \patchcmd{\titleblock@produce}
  {\frontmatter@RRAPformat}
  {\frontmatter@RRAPformat{\produce@RRAP{*#1\href{mailto:#2}{#2}}}\frontmatter@RRAPformat}
  {}{}
}%
\makeatother

\begin{document}

\preprint{AIP/123-QED}

\title{Ultrafast electron diffraction of photoexcited gas-phase cyclobutanone predicted by {\it ab initio} multiple cloning simulations}


\author{Dmitry V.  Makhov} \email{d.makhov@leeds.ac.uk}
\affiliation{\leeds}
\affiliation{\bristol}
\author{Adam Kirrander}
\affiliation{\oxchem}
\author{Dmitrii V. Shalashilin}
\affiliation{\leeds}

\date{\today}

\begin{abstract}
We present the result of our calculations of ultrafast electron diffraction (UED) for cyclobutanone excited into  $S_2$ electronic state, which are based on the non-adiabatic dynamics simulations with \textit{Ab Initio} Multiple Cloning (AIMC) method with the electronic structure calculated at the SA(3)-CASSCF(12,12)/aug-cc-pVDZ level of theory. The key features in the UED pattern were identified that can be used to distinguish between the reaction pathways observed in the AIMC dynamics, although there is a significant overlap between representative signals due to structural similarity of the products. The calculated UED pattern can be compared with experiment.
\end{abstract}

\maketitle

\section{\label{sec:introduction} Introduction}

\par Ultrafast electron diffraction (UED) has evolved into a powerful method for structural dynamics.\cite{CenturionRMP2022,MillerCR2017} Although UED, and the closely related method of ultrafast x-ray scattering,\cite{YongCh9Kasra,Stankus2020review} arguably provide the most direct access to structural dynamics in photoexcited molecules, the interpretation of experiments is nontrivial. Despite significant progress in the development of inverse methods, which aim to produce a (time-dependent) molecular model commensurate with the experimental data,\cite{yang_reconstruction_2014,ishikawa_direct_2015,Asenov2020,YongFD2021,Acheson2023inversion} the gold standard for interpreting ultrafast experiments remains comparison to high-quality simulations of the photoexcited target molecule. However, such simulations remain challenging, and their veracity depends keenly on numerous methodological choices. As a much needed step towards surveying and evaluating good practice, the Journal of Chemical Physics recently announced the {\it{Prediction Challenge: Cyclobutanone Photochemistry}} to which this paper is a response.

\par The challenge is motivated by an experiment at the SLAC Megaelectronvolt ultrafast electron diffraction (SLAC MeV-UED) facility, where a gas-phase sample of cyclobutanone is irradiated with a 200 nm laser pulse and a time-resolved UED signals are recorded. At this excitation energy, a low-lying n$\rightarrow$3s (S$_2$) Rydberg state in cyclobutanone is excited.\cite{Drury-Lessard1978, OToole1991, Whitlock1971,Kuhlman2012} The photochemistry of cyclobutanone has a long history, with early experiments carried out in the 1940s,\cite{Benson1942} with particular attention having been paid to the photoproducts of the reaction.\cite{Benson1942,Denschlag1968,Lee1969,Lee1971,Diau2001,Kao2020}

\par In this work, the photodynamics of cyclobutanone is simulated using \textit{Ab initio} Multiple cloning (AIMC) \cite{AIMC, MAKHOV2017200, C8CP02321B} approach, which is in principle a fully quantum, formally exact methodology based on using Gaussian Coherent States propagated by Ehrenfest trajectories as a basis for quantum dynamics of nuclear wave functions. AIMC was successfully applied before \cite{Pyrrole,2EP, pyrazole, Vib, Vib_D} to simulate the process of the photodissociation of a number of heterocyclic molecules. Based on AIMC dynamics results, the gas phase time-resolved UED pattern of cyclobutanone photoexcited using a 200 nm pulse is calculated for the initial 200 fs of dynamics, allowing for direct comparison to experimental data.

\section{\label{sec:theory}Theoretical methods}

\subsection{AIMC}

\par As the AIMC methodology was extensively described before \cite{AIMC, MAKHOV2017200, C8CP02321B, Pyrrole}, here we provide only a summary of the technique. The AIMC method represents the further development of  the Multi Configurational Ehrenfest (MCE) \cite{DS2009, DS2010, DS2011, Saita} approach and makes use of the following wave-function ansatz:
	\begin{equation} 
		|\Psi(\textbf{R},\textbf{r},t)\rangle=\sum_n c_n(t)|\chi(\textbf{R},t)\rangle\sum_I a_I^{(n)}|\phi_I(\textbf{r};\textbf{R})\rangle, \label{wavefunct-tot}
	\end{equation}
where \textbf{R} and \textbf{r} are electronic and nuclear coordinates respectively. The electronic part of each basis function is represented in a basis of are adiabatic electronic states $|\phi_I \rangle$, and the nuclear parts is a moving Gaussian Coherent State:

\[|\chi_n({\bf R},t)\rangle=\left(\frac{2\alpha}{\pi}\right)^{\frac{N_{\text{dof}}}{4}}\times\]
	\begin{equation}
		\exp\left(-\alpha\left({\bf R}-{\bf R}_n\right)^2+\frac{i}{\hbar}{\bf P}_n\left({\bf R}-{\bf R}_n\right)+\frac{i}{\hbar}\gamma_n(t)\right)\label{CS},
	\end{equation}
 which is a Gaussian-shaped de Broglie wave centred at $\textbf{R}_n$ with momentum $\textbf{P}_n$ and phase $\gamma_n$. The motion of Gaussians $|\chi_n \rangle$ is guided by the Ehrenfest force:

	\begin{equation*}
		\dot {\bf R}_n=M^{(-1)}{\bf P}_n,
	\end{equation*}
 \begin{multline}
		{\dot {\bf P}_n}=-\sum_I\left|a_I^{(n)}\right|^2\nabla V_I^{(n)} \\ +\sum_{I\neq J} a_I^{(n)*}a_J^{(n)}{\bf d}_{IJ}^{(n)}\cdot\dot {\bf R}_n\left(V_I^{(n)}-V_J^{(n)}\right),\label{force} 
	\end{multline}
	where $V_I$ is potential energy surface of the \textit{I}th electronic state, ${\bf d}_{IJ}$ a non-adiabatic coupling vector, and $M$ is a diagonal matrix of atomic masses. As the force depends on Ehrenfest amplitudes $a_I^{(n)}$, the equations of motion (\ref{force}) must be solved simultaneously with the equations for $a_I^{(n)}$: 

	\begin{equation}
		\dot a_I^{(n)}=-\frac{i}{\hbar}V_I^{(n)}a_I^{(n)}-\sum_J\dot {\bf R}_n\cdot {\bf d}_{IJ}a_J^{(n)},\label{el_Sch}
	\end{equation}
 where the right side is the electronic Hamiltonian for \textit{n}th basis function. Finally, phase $\gamma_n$ is propagated semiclassically as $\dot \gamma_n={\bf P}_n \dot {\bf R}_n/2$. 
 
	It is well known that Ehrenfest trajectories misguide basis set when several non-interacting electronic states have significant amplitudes. The cloning procedure is applied in AIMC approach in order to address this issue. In principle, the cloning can be viewed as straightforward way of spawning employed in the Multiple Spawning method \cite{Curchod2018d}. The idea of cloning is to replace a basis function with two clones, each of which is guided in most by just one potential energy surface. In the simplest case of two electronic states: 
	\begin{multline}
c_n |\chi_n\rangle \left(a_1^{(n)} |\phi_1\rangle + a_2^{(n)} |\phi_2\rangle\right) = \\ c_n' |\chi_n\rangle \left(0 \times |\phi_1\rangle + \frac {a_2^{(n)}}{|a_2^{(n)}|} |\phi_2\rangle\right) \\ + c_n'' |\chi_n\rangle \left(\frac {a_1^{(n)}}{|a_1^{(n)}|} |\phi_1\rangle + 0 \times |\phi_2\rangle \right),
	\end{multline}
where
\begin{equation}
c_n'=c_n |a_2^{(n)}|; c_n''=c_n |a_1^{(n)}|.
\end{equation}
The total contribution of two clones into wave function $|\Psi(\textbf{R},\textbf{r},t)\rangle$ is exactly the same as that of the original basis function. However, cloning increases the size of basis set creating additional flexibility, as two clones can now move in different directions. The cloning is applied when the breaking force  $\textbf{F}_I^{(br)} = |a_I|^2  \left( \nabla V_I - \sum_J |a_J|^2 \nabla V_J \right)$ exceeds a threshold and, at the same time, the magnitude of non-adiabatic coupling is below a second threshold. Cloning is an extremely important part of AIMC method, as it allows AIMC to reproduce the bifurcation of the wave function at conical intersections.

The trajectories in AIMC approach can be calculated independently using potential energies forces and non-adiabatic coupling vectors calculated “on the fly” by an electronic structure code. Then, time-depended Schrödinger equation for amplitudes $c_n$ is solved in post-processing in the precalculated trajectory-guided basis (\ref{wavefunct-tot}). 

In practice to achieve good convergence, a number of sampling techniques have to be used. Swarms of coupled trajectory guided Gaussians, as well as their train guided by the same trajectories are among those techniques \cite{DS2008}. It has been demonstrated that MCE can produced the results, which are well converged \cite{EFFCTSMPLNG} and AIMC, its \textit{ab initio} direct dynamics version, is more accurate than Surface Hopping or Ehrenfest Dynamics \cite{CmprsnCnvrg, NEXMD}. A technique which allows to take into account pulse shape and dynamics which occur during the excitation has been developed \cite{MAKHOV201846}.  In its simplest form, which is used in the present paper, AIMC can yield qualitative or semiquantitative picture of the process, similar to that given by Surface Hoping, \textit{Ab initio} Multiple Spawning (AIMS) \cite{Curchod2018d} and many other popular techniques.

\subsection{Ultrafast electron diffraction\label{sec:IAM}}

For the modelling of the UED signals, we anticipate that the experimental data in this challenge can be modelled reliably using the independent atom model (IAM). This approximates the scattering signal as a coherent sum of scattering from isolated atoms centered at the positions of the nuclei in the target molecule. Notably, this model excludes the contribution of the bonding electrons and the characteristics of the electronic states of the molecule. Should the quality of the experimental data necessitate that these effects are accounted for, then numerical codes capable of this exist, \cite{moreno_carrascosa_ab_2019,Parrish2019,zotev_excited_2020,Moreno2022} albeit at significantly higher computational cost.

The total (energy-integrated) scattering cross section into the solid angle $d\Omega$ at time $t$ is given by,\cite{kirrander_ultrafast_2016,Minas2017,kirrander_fundamental_2017}
\begin{equation} \label{eq:cross-section}
    \frac{d\sigma}{d\Omega} \bigg/  \left(\frac{d\sigma}{d\Omega}\right)_{\mathrm{Rh}} =  I_\mathrm{tot}(\mathbf{s},t),
\end{equation}
where $\mathbf{s}=\mathbf{k}_0-\mathbf{k}_1$ is the scattering vector expressed in terms of the wave vectors of the incoming and outgoing electrons. The scattering is given in units of the Rutherford cross section $(d\sigma/d\Omega)_{\mathrm{Rh}}$ which includes the $s^{-4}$ scaling factor.\cite{mott_scattering_1930, bethe_zur_1930} Note that the expression above does not account for the duration of the electron pulse, which may be accounted for via a temporal convolution of the predicted signal by the instrument response function for the experiment.

General expressions that account for the full wavefunction in Eq.~(\ref{wavefunct-tot}), including the non-local nature of the individual Gaussian coherent states, have been derived previously.\cite{kirrander_ultrafast_2016} Given the sparse basis used in the present simulations, we resort here to the diagonal bracket-averaged Taylor (BAT) expansion approximation\cite{kirrander_ultrafast_2016} and assume that expansion coefficients are independent of time, $c_n \approx c_n(t)$, giving the total scattering intensity as,
\begin{equation}
    I_{\mathrm{tot}}\left(\mathbf{s},t\right) = \sum_{n=1} |c_n|^2 I_n\left(\mathbf{s}, \mathbf{R}_n(t)\right).
\end{equation}
In this simplified form, sufficient for our present needs, the scattering from each trajectory is given by IAM as,\cite{SimmermacherCh3Kasra}
\begin{equation} \label{eq:IAM-trj}
   I_n(\mathbf{s}, \mathbf{R}_n(t)) = |F(\mathbf{s},\mathbf{R}_n(t))|^2 + S_\mathrm{inel}(s), 
\end{equation}
where $S_\mathrm{inel}(s)$ is the inelastic scattering, which is independent of molecular geometry and isotropic, as underscored by its dependence only on the amplitude of the momentum transfer vector, $s=|\mathbf{s}|$. It is given by an incoherent summation over the individual atomic contributions,   
\begin{equation}
    S_\mathrm{inel}(s) = \sum_{A=1}^{N_\mathrm{at}} S_A(s),
\end{equation}
with $N_\mathrm{at}$ the number of atoms in the molecule. The corresponding elastic contribution is given by the form factor $F(\mathbf{s},\mathbf{R}_n(t))$,
\begin{equation} \label{eq:IAM}
    F(\mathbf{s},\mathbf{R}_n(t)) = \sum_{A=1}^{N_\mathrm{at}} f^e_A(s) e^{\imath \mathbf{s} \mathbf{R}_{nA}(t)},
\end{equation}
where $f^\mathrm{e}_A(s)$ are the atomic form factors and $\mathbf{R}_{nA}(t)$ the position vector for atom $A$ in trajectory $n$. The form factors for electron scattering are $f_{A}^\mathrm{e}=\left(f_{A}^\mathrm{x}-Z_A\right)$, where $f_A^{\mathrm{x}}$ the x-ray scattering form factor and $Z_A$ is the atomic number.\cite{mott_scattering_1930, bethe_zur_1930} Both $f_{A}^\mathrm{x}(s)$ and $S_A(s)$ are tabulated.\cite{IntTabCryVolC} For high energy electron scattering it is sometimes necessary to use form factors with relativistic corrections,\cite{salvat_elastic_1991, salvat_elsepadirac_2005} but this is not done presently.

When the target is a gas of anisotropic molecules, $|F(\mathbf{s},\mathbf{R}_n(t))|^2$ in Eq.\ (\ref{eq:IAM-trj}) is replaced by its rotationally averaged counterpart, $\langle |F(s,\mathbf{R}_n(t))|^2 \rangle$,\cite{debye_zerstreuung_1915}
\begin{equation}\label{eq:IAM_rotav}
        \langle |F(s,\mathbf{R}_n(t))|^2 \rangle = \sum_{A,B}^{N_\mathrm{at}} f_{A}^\mathrm{e}\left(s\right) f_{B}^\mathrm{e}\left(s\right) \frac{\sin{\left(s R_{nAB}(t) \right)}}{s R_{nAB}(t)}
\end{equation}
where $R_{nAB}(t) = |\mathbf{R}_{nA}(t) - \mathbf{R}_{nB}(t)|$ is the distance between atoms $A$ and $B$ in trajectory $n$.

\section{\label{Methods}Computational details}

The trajectories were calculated using our own AIMC code,  where potential energies, forces and non-adiabatic couplings given “on the fly” by MOLPRO \cite{Molpro} electronic structure package at  SA(3)-CASSCF(12,12)/aug-cc-pVDZ level of theory. We note that the electronic structure method has been benchmarked in another paper submitted to the same challenge by one of the co-authors (AK). In brief, three electronic states were taken into consideration, a ground state and two lowest singlet excited states. Higher energy Rydberg states have been shown to exist (3p character) but have been shown to be unimportant for dynamics after excitation into S$_2$, therefore we do not include them in our simulations. \cite{Kuhlman2012a} We also do not take triplet states into consideration in this work, as they have been shown to only play a role in dynamics upon excitation with long wavelengths. \cite{Kao2020} The initial positions and momenta for all trajectories are randomly sampled from the ground state vibrational Wigner distribution using vibrational frequencies and normal modes calculated at the same level of theory. This ground state wavepacket is then simply lifted to to the second excited state within Condon approximation. As in our previous simulations \cite{AIMC,Pyrrole,2EP, pyrazole, Vib, Vib_D} the cloning thresholds were taken as $5\cdot 10^{-6}$ a.u.\ and $2\cdot 10^{-3}$ a.u.\ for the magnitude of breaking force and non-adiabatic coupling, respectively.

In {\it ab initio} dynamics, the number of trajectories is severely limited by the high cost of electronic structure calculations (especially for larger molecules). When the initial multi-dimensional wave-function is randomly sampled with a small number of Gaussians, these Gaussians will be located far away from each other with no coupling between them. Running Gaussians closer together would be an inefficient use of CPU time unless we need to reproduce a particular quantum effect of the nuclear motion. In this work, we use a simplified semiclassical version of AIMC, where we do not consider the coupling between the trajectories. Instead, each branch simply gets its amplitudes at the time of cloning, and this amplitude determines the statistical weight of that brunch. 

We initially run an ensemble of 39 Efrenfest trajectories, which give rise to 121 branches in the process of cloning. All trajectories were propagated for 200 fs with ~0.06 fs (2.5 a.u.) timestep. A relatively small number of trajectories and short duration of the dynamics is due to the strict deadline for this work. Nevertheless, despite not very good statistics, our calculations show clear UED patterns for the cyclobutanone photodynamics.

\section{\label{Results}Results and discussion}

\subsection{\label{sec:dynamics}Dynamics}

\begin{figure}[h]
			\centering
			\includegraphics[width=0.5\textwidth]{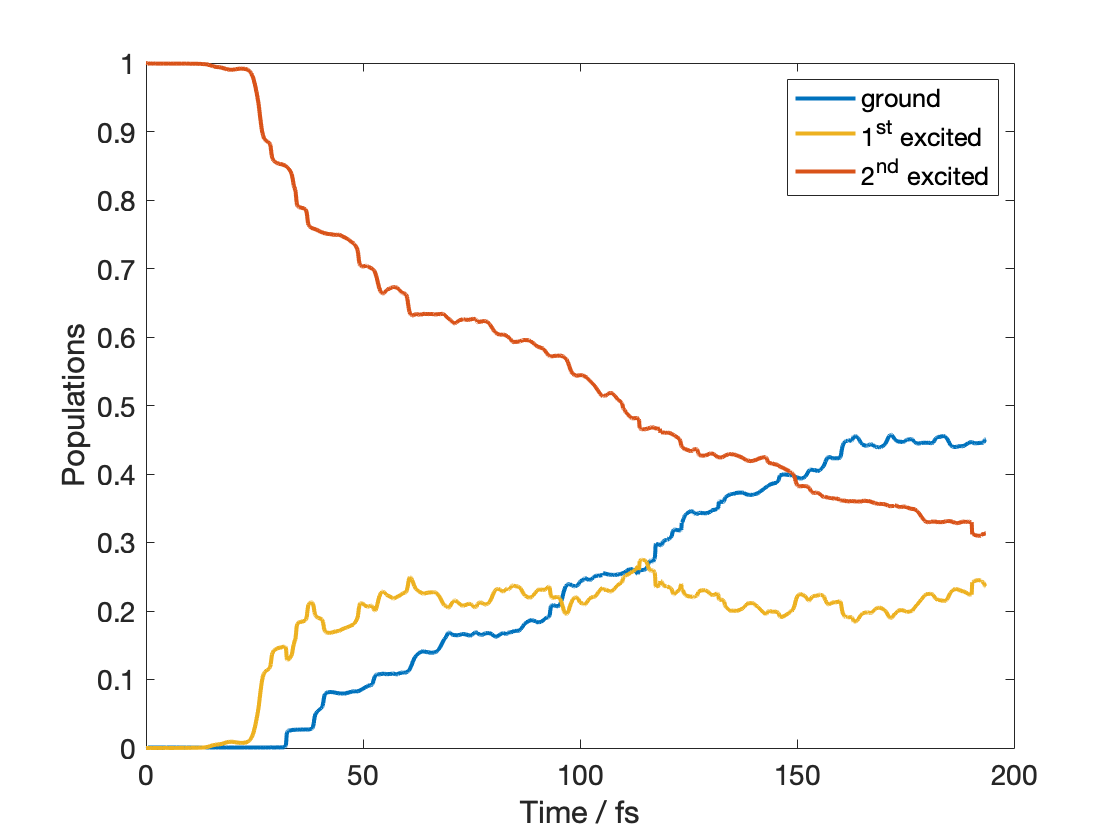}
			\caption{The dynamics of $S_2$ (red), $S_1$ (yellow) and $S_0$ (blue) electronic state populations for cyclobutanone molecule after its photoexcitation into $S_2$ state.}\label{Populations}
		\end{figure}

Figure \ref{Populations} presents the dynamics of the populations for $S_0$, $S_1$ and $S_2$ electronic states. For the first 25 fs of dynamics, the molecules stay in $S_2$  state, then  $S_2 \rightarrow S_1$ population transfer starts. The growing population of $S_1$  state immediately initiate the next step of population transfer, from $S_1$ into the ground state. Within next 10 fs, the $S_1$ state population reaches the  equilibrium level of about 20\%, when  the rates of  $S_2 \rightarrow S_1$ and  $S_1 \rightarrow S_0$ transfers are about the same. For the rest of our dynamics, the $S_1$ state population is fluctuating around this level, while $S_2$  state exhibit the exponential decay into a ground state $S_2 \rightarrow S_1 \rightarrow S_0$. 

\begin{figure}[h!]
			\centering
			\includegraphics[width=0.5\textwidth]{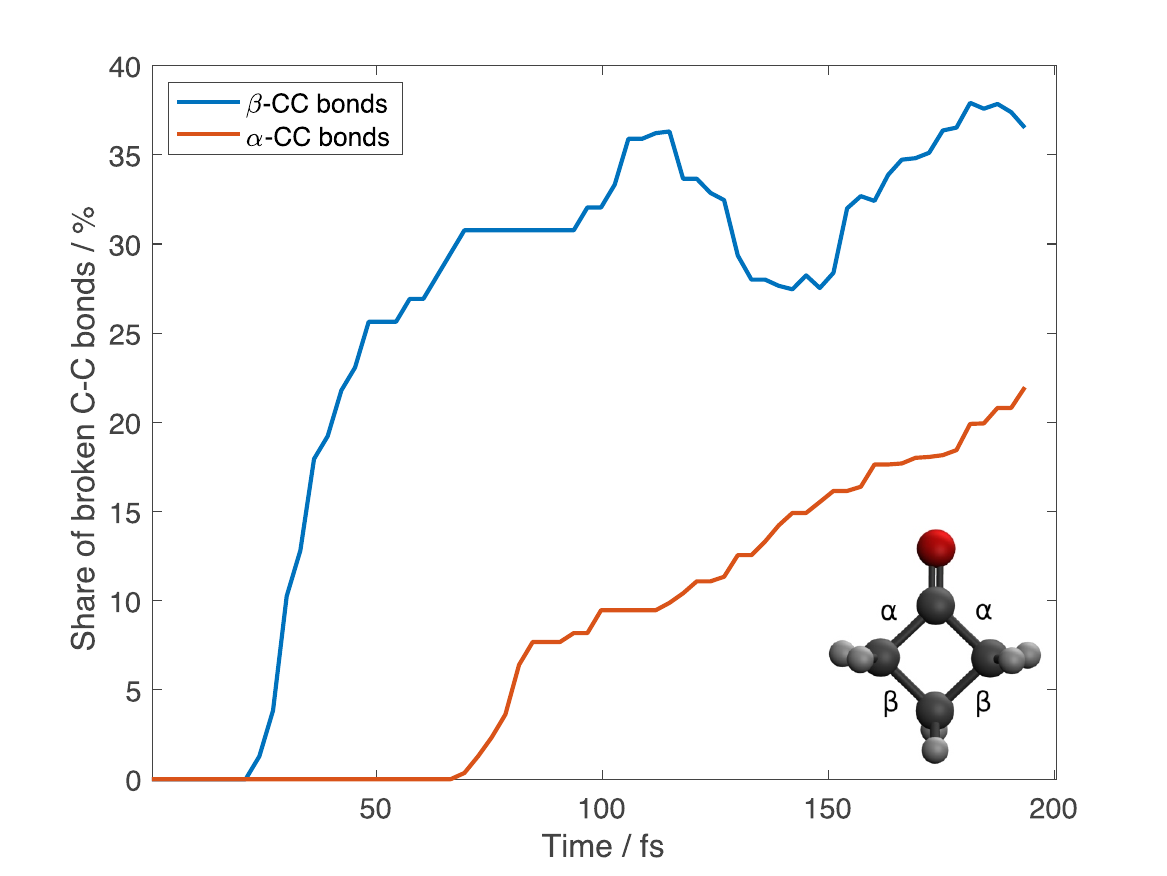}
			\caption{The share of broken $\mathrm{\alpha}$- (red) and  $\mathrm{\beta}$- (blue) CC bonds as a function of time for cyclobutanone molecule after its photoexcitation into $S_2$ state.}\label{Bonds}
		\end{figure}

Figure \ref{Bonds} shows the dynamics of C-C bonds breaking in the cyclobutanone ring. The bond is considered broken when the distance between two atoms exceeds 3 \AA. The process of ring opening starts at about 25 fs time, simultaneously with the beginning of the non-adiabatic decay of $S_2$ state, by breaking  $\mathrm{\beta}$-CC bonds. Within next 50 fs, 30\% of these bonds break, which corresponds to 60\% of the opened rings.  At this stage of the dynamics, $\mathrm{\alpha}$-CC bonds are starting to break; this happens mostly in already opened rings creating ethylene \ce{(CH2)(CH2)} and ethenone \ce{(CO)(CH2)} molecules. In some of these ethenone molecules, C=C bonds also later break, creating CO and CH2 radicals. 

After about 100 fs time, some opened rings are beginning to close again (or, at least, their ends approach each other to less than 3 \AA). Later, the ring can open again, creating an oscillatory behaviour in the number of broken bonds. 

By the end of the dynamics, the yield in the \ce{(CH2)(CH2)} + \ce{(CO)(CH2)} dissociation channel is 40.6\% , the yield in the \ce{(CH2)(CH2)} + \ce{(CO)} + \ce{(CH2)} dissociation channel is 3.5\%, and the yield of ring opening is 31.0\%; also 17.2\% molecules have remained in the closed ring form. The remaining 7.7\% are found at various other intermediate configurations at the end of our 200 fs dynamics; the longer-term dynamics will be a subject of our future work.

\subsection{Ultrafast electron diffraction}

\par
\begin{figure}
    \centering
    \includegraphics[width=0.49\textwidth]{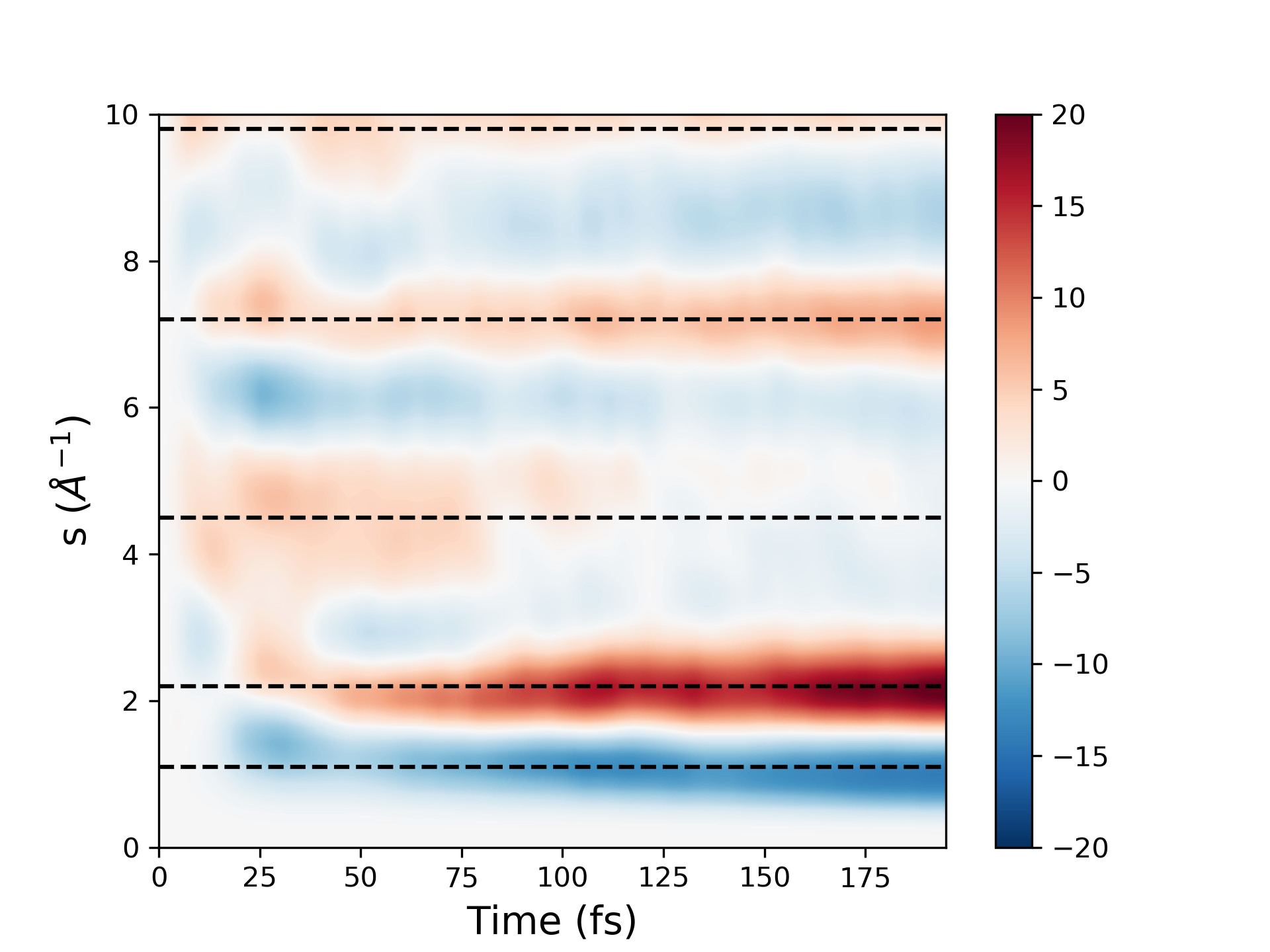}
    \caption{Gas-phase UED pattern, $\%I_\mathrm{tot}(s,t)$ in Eq.\ (\ref{eq:percI}), for cyclobutanone calculated using the IAM with all trajectories and branches from our dynamics simulations. Five key features in the UED pattern are highlighted with horizontal dashed lines.}
    \label{fig:ued_full}
\end{figure}

 The AIMC simulations presented in section \ref{sec:dynamics} serve as a framework to calculate the total rotationally averaged UED pattern for cyclobutanone, using the methodology presented in Section \ref{sec:IAM}. The UED signal thus obtained is given in Fig.\ \ref{fig:ued_full}, plotted as percent difference $\%I_\mathrm{tot}(s,t)$,
\begin{equation} \label{eq:percI}
    \%I_\mathrm{tot}(s,t) = 100 \times \frac{I_\mathrm{tot}(s,t) - I_\mathrm{tot}(s,0)}{I_\mathrm{tot}(s,0)},
\end{equation}
\noindent where, $I_\mathrm{tot}(s,t)$ is the signal at time $t$ and $I_\mathrm{tot}(s,0)$ is the reference signal at $t=0$, i.e.\ the pump-off signal.

\par
\begin{figure*}
    \centering
    \includegraphics[width=\textwidth]{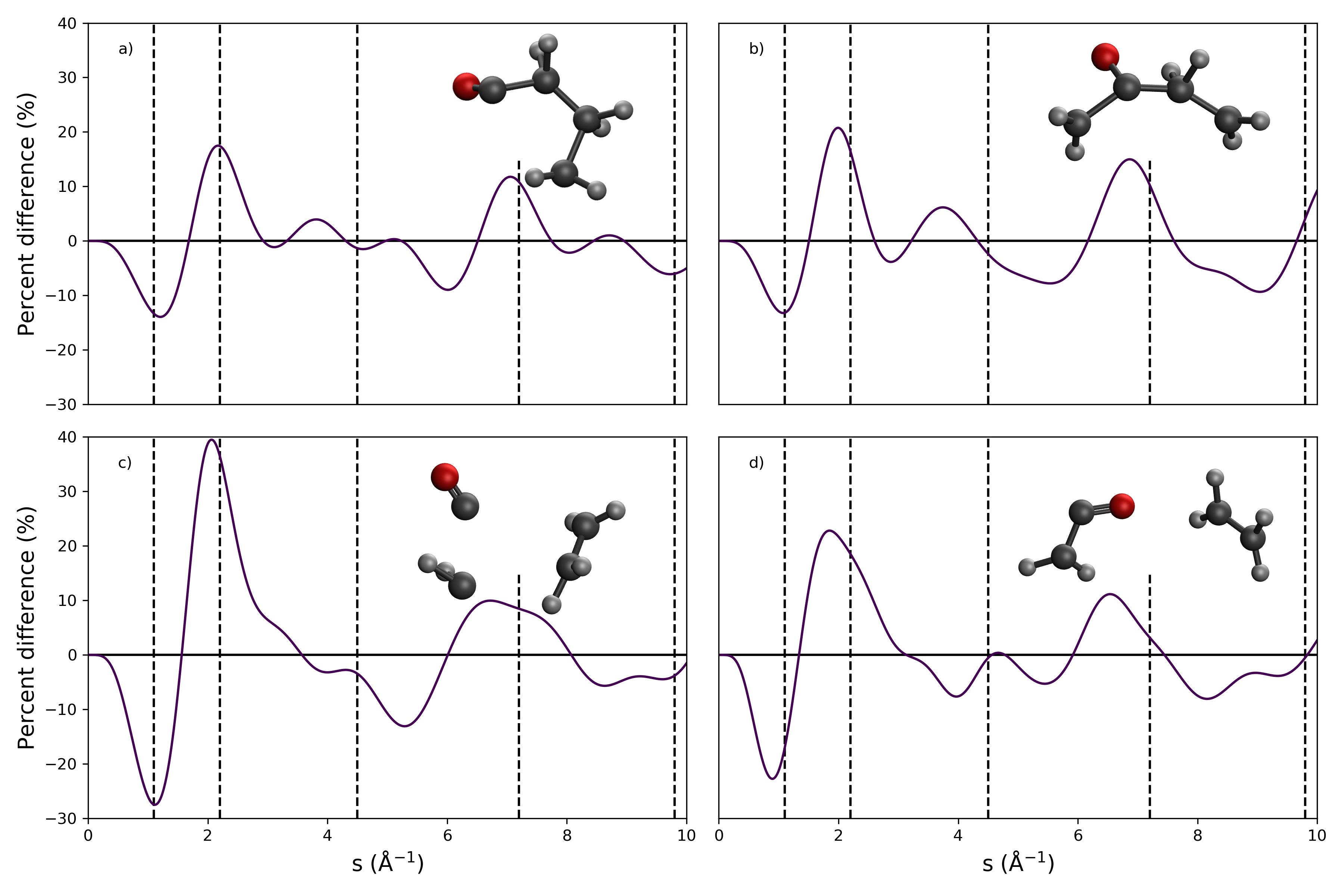}
    \caption{Static signals obtained for representative structures of pathways observed in AIMC simulations given in percent differences. Panel a) $\mathrm{\alpha}$-CC ring opening, b) $\mathrm{\beta}$-CC ring opening, c) dissociation to form \ce{CO}, \ce{C_2H_4} and \ce{CH_2}, and finally d) \ce{CH_2CO} and \ce{C_2H_4}. The structures shown as insets are the geometries from which the static signal is calculated. The five features shown in Figure \ref{fig:ued_full} are highlighted with vertical dashed lines. }
    \label{fig:static_signals}
\end{figure*}

 To aid the interpretation of Fig.\ \ref{fig:ued_full}, we have also calculated the static signal for all reaction products observed in the AIMC simulations, shown in Fig.\ \ref{fig:static_signals}. Representative structures were taken from the trajectories showing each reaction observed and the IAM was then used to calculate the UED signal for each structure, briefly comprising of: a) $\mathrm{\alpha}$-CC bond breaking, b) $\mathrm{\beta}$-CC bond breaking, c) the production of \ce{C_2H_4}, \ce{CH_2} and \ce{CO}, and also d) production of \ce{CH_2CO} and \ce{C_2H_4}. The structures for these pathways can be seen in the insets of Fig.\ \ref{fig:static_signals}.

\par Five key features can be observed in the gas-phase UED of cyclobutanone, these are highlighted by horizontal dashed lines in Fig.\ \ref{fig:ued_full} at $s= 1.1,\; 2.2,\; 4.5,\; 7.2$ and $9.8$ \AA$^{-1}$. Matching these peaks with those observed in Fig.\ \ref{fig:static_signals}, one can see all four reaction products yield a negative feature at $\sim$1.1 \AA$^{-1}$ and a positive feature at $\sim$2.2 \AA$^{-1}$. Both features at $\sim$1.1 and $\sim$2.2 \AA$^{-1}$ grow in intensity after approximately 20 fs, matching the timescale shown in Section \ref{sec:dynamics}. Both features continue to grow in intensity. Most notably, around 75 fs, Fig.\ \ref{Bonds} shows that we begin to observe $\mathrm{\alpha}$-CC bond breaking. This, coupled with the high intensity observed in the static signal of \ref{fig:static_signals}c) at $2.2$ \AA$^{-1}$, suggest that the stepwise mechanism to form \ce{CO}, \ce{C_2H_4} and \ce{CH_2} requires approximately 75 fs to form these products. However, we must note that the features in the signal have contributions from all the pathways shown in Fig.\ \ref{fig:static_signals}.

\par An additional broad positive feature can be observed centered at $4.5$ \AA$^{-1}$, with a signal that decays after approximately 115 fs. Figure \ref{fig:static_signals}c) shows a broad negative feature between the values $\sim$3.75 \AA$^{-1}$ and $\sim$5.5 \AA$^{-1}$, indicating this reaction pathway causes the depletion of the broad signal. Once again due to the structural similarities of other reaction pathways, the net signal is a compounded signal with contributions from all products. A similar depletion can be seen in Fig.\ \ref{fig:static_signals}d), due to the higher proportion of trajectories being classified as belonging to d) this likely has a stronger effect on the signal. 

\par Further peaks can be seen at 7.5 and 9.8 \AA$^{-1}$. A peak at 7.5 \AA$^{-1}$ can be seen in all pathways with a similar intensity ($\sim$10\%) therefore yielding little structural information other than the molecule has moved away from the equilibrium geometry. In contrast, Fig.\ \ref{fig:static_signals}c) and to a lesser extent Fig.\ \ref{fig:static_signals}d) both show a signal at 9.8 \AA$^{-1}$. Thus, it is likely this feature arises from the breaking of the $\mathrm{\beta}$-CC common to both reaction products.

\section{Conclusions \label{sec:conclusion}}

\par This work was undertaken in response to the {\it{“Prediction Challenge: Cyclobutanone Photochemistry”}} and presents simulated ultrafast electron diffraction (UED) signals for gas-phase cyclobutanone upon photoexcitation into the $S_2$ electronic state. The main dissociation pathways of photoexcited cyclobutanone were identified with the help of AIMC non-adiabatic dynamics. Then, using these AIMC trajectories, the electronic diffraction was calculated using the IAM method. The calculated UED pattern was compared with static signals for representative structures for the different dissociation pathways observed in the AIMC dynamics. Overall, five key features in the UED pattern can be used to distinguish the reaction channels observed in the AIMC simulation. We find that there is significant overlap between many features due to a high degree of structural similarity between the different photoproducts, combined with a significant degree of symmetry in cyclobutanone. However, ultimately, we found strong correlations between the timescales and products evident in the simulations and features in the overall UED signal (shown in Fig.\ \ref{fig:ued_full}). 

\par The extent of the work was limited by the strict deadline inherent in the {\it{Challenge}}, and it is therefore straightforward to identify avenues for further work. To begin with, the presented simulations include only the first 200 fs of the cyclobutanone dissociation dynamics and the degree of sampling, i.e. the number of trajectories propagated, was also limited.  Previously we developed a technique which allows to include the pump pulse shape into account and allows to account for the dynamics during the pulse \cite{MAKHOV201846}. We have not used this approach here and assumed instant excitation, but it can be straightforwardly done. With longer propagation times it will also be easy to account for some coupling between coherent states, using the so called train basis functions \cite{MAKHOV2017200}. This approach does not require additional trajectories and electronic structure calculations.  All of these improvements will be the subject of subsequent work. Longer simulation times will make it possible to make comparisons to the long-term dynamics observed in the experiment, while more trajectories should in principle allow us to move beyond the independent and semiclassical trajectory approximation used when calculating the UED signals. Also, as discussed in the UED theory section, should the experimental data indicate that more subtle effects in the scattering were observed, then {\it{ab initio}} simulations of the scattering signal, going beyond the independent atom model, are clearly of interest.

\par In summary, the present work demonstrates the capability of AIMC to simulate photodynamics in a challenging molecule and that UED signals can be predicted straightforwardly from the simulations. We also note that the AIMC simulations should in principle provide a better basis for the prediction of experimental signals that reflect the degree of coherence in the molecule during the dynamics, such as nonlinear spectroscopies or coherent mixed scattering.\cite{simmermacher_theory_2019,Keefer2021}

\section*{Acknowledgement}
DM and DS acknowledge the support of EPSRC grant EP/P021123/1. AK  acknowledges funding from the Leverhulme Trust (RPG-2020-208), and EPSRC grant EP/V049240.

\bibliography{CYCLOBUTANONE}

\begin{thebibliography}{55}%
\makeatletter
\providecommand \@ifxundefined [1]{%
 \@ifx{#1\undefined}
}%
\providecommand \@ifnum [1]{%
 \ifnum #1\expandafter \@firstoftwo
 \else \expandafter \@secondoftwo
 \fi
}%
\providecommand \@ifx [1]{%
 \ifx #1\expandafter \@firstoftwo
 \else \expandafter \@secondoftwo
 \fi
}%
\providecommand \natexlab [1]{#1}%
\providecommand \enquote  [1]{``#1''}%
\providecommand \bibnamefont  [1]{#1}%
\providecommand \bibfnamefont [1]{#1}%
\providecommand \citenamefont [1]{#1}%
\providecommand \href@noop [0]{\@secondoftwo}%
\providecommand \href [0]{\begingroup \@sanitize@url \@href}%
\providecommand \@href[1]{\@@startlink{#1}\@@href}%
\providecommand \@@href[1]{\endgroup#1\@@endlink}%
\providecommand \@sanitize@url [0]{\catcode `\\12\catcode `\$12\catcode `\&12\catcode `\#12\catcode `\^12\catcode `\_12\catcode `\%12\relax}%
\providecommand \@@startlink[1]{}%
\providecommand \@@endlink[0]{}%
\providecommand \url  [0]{\begingroup\@sanitize@url \@url }%
\providecommand \@url [1]{\endgroup\@href {#1}{\urlprefix }}%
\providecommand \urlprefix  [0]{URL }%
\providecommand \Eprint [0]{\href }%
\providecommand \doibase [0]{http://dx.doi.org/}%
\providecommand \selectlanguage [0]{\@gobble}%
\providecommand \bibinfo  [0]{\@secondoftwo}%
\providecommand \bibfield  [0]{\@secondoftwo}%
\providecommand \translation [1]{[#1]}%
\providecommand \BibitemOpen [0]{}%
\providecommand \bibitemStop [0]{}%
\providecommand \bibitemNoStop [0]{.\EOS\space}%
\providecommand \EOS [0]{\spacefactor3000\relax}%
\providecommand \BibitemShut  [1]{\csname bibitem#1\endcsname}%
\let\auto@bib@innerbib\@empty
\bibitem [{\citenamefont {Filippetto}\ \emph {et~al.}(2022)\citenamefont {Filippetto}, \citenamefont {Musumeci}, \citenamefont {Li}, \citenamefont {Siwick}, \citenamefont {Otto}, \citenamefont {Centurion},\ and\ \citenamefont {Nunes}}]{CenturionRMP2022}%
  \BibitemOpen
  \bibfield  {author} {\bibinfo {author} {\bibfnamefont {D.}~\bibnamefont {Filippetto}}, \bibinfo {author} {\bibfnamefont {P.}~\bibnamefont {Musumeci}}, \bibinfo {author} {\bibfnamefont {R.}~\bibnamefont {Li}}, \bibinfo {author} {\bibfnamefont {B.}~\bibnamefont {Siwick}}, \bibinfo {author} {\bibfnamefont {M.}~\bibnamefont {Otto}}, \bibinfo {author} {\bibfnamefont {M.}~\bibnamefont {Centurion}}, \ and\ \bibinfo {author} {\bibfnamefont {J.}~\bibnamefont {Nunes}},\ }\bibfield  {title} {\enquote {\bibinfo {title} {Ultrafast electron diffraction: {Visualizing} dynamic states of matter},}\ }\href {\doibase 10.1103/RevModPhys.94.045004} {\bibfield  {journal} {\bibinfo  {journal} {Rev. Mod. Phys.}\ }\textbf {\bibinfo {volume} {94}},\ \bibinfo {pages} {045004} (\bibinfo {year} {2022})}\BibitemShut {NoStop}%
\bibitem [{\citenamefont {Ischenko}, \citenamefont {Weber},\ and\ \citenamefont {Miller}(2017)}]{MillerCR2017}%
  \BibitemOpen
  \bibfield  {author} {\bibinfo {author} {\bibfnamefont {A.~A.}\ \bibnamefont {Ischenko}}, \bibinfo {author} {\bibfnamefont {P.~M.}\ \bibnamefont {Weber}}, \ and\ \bibinfo {author} {\bibfnamefont {R.~J.~D.}\ \bibnamefont {Miller}},\ }\bibfield  {title} {\enquote {\bibinfo {title} {Capturing chemistry in action with electrons: Realization of atomically resolved reaction dynamics},}\ }\href {\doibase 10.1021/acs.chemrev.6b00770} {\bibfield  {journal} {\bibinfo  {journal} {Chemical Reviews}\ }\textbf {\bibinfo {volume} {117}},\ \bibinfo {pages} {11066--11124} (\bibinfo {year} {2017})}\BibitemShut {NoStop}%
\bibitem [{\citenamefont {Yong}, \citenamefont {Kirrander},\ and\ \citenamefont {Weber}(2024)}]{YongCh9Kasra}%
  \BibitemOpen
  \bibfield  {author} {\bibinfo {author} {\bibfnamefont {H.}~\bibnamefont {Yong}}, \bibinfo {author} {\bibfnamefont {A.}~\bibnamefont {Kirrander}}, \ and\ \bibinfo {author} {\bibfnamefont {P.~M.}\ \bibnamefont {Weber}},\ }\bibfield  {title} {\enquote {\bibinfo {title} {Time-resolved x-ray scattering of excited state structure and dynamics},}\ }in\ \href {\doibase https://doi.org/10.1039/9781837671564} {\emph {\bibinfo {booktitle} {{Structural Dynamics with X-ray and Electron Scattering}}}},\ \bibinfo {series} {{Theoretical and Computational Chemistry Series}}, Vol.~\bibinfo {volume} {25},\ \bibinfo {editor} {edited by\ \bibinfo {editor} {\bibfnamefont {K.}~\bibnamefont {Amini}}, \bibinfo {editor} {\bibfnamefont {A.}~\bibnamefont {Rouz\'{e}e}}, \ and\ \bibinfo {editor} {\bibfnamefont {M.~J.~J.}\ \bibnamefont {Vrakking}}}\ (\bibinfo  {publisher} {Royal Society of Chemistry},\ \bibinfo {address} {United Kingdom},\ \bibinfo {year} {2024})\ \bibinfo {edition} {1st}\ ed.,\ Chap.~\bibinfo {chapter} {3}, p.\ \bibinfo
  {pages} {344},\ \bibinfo {note} {www.rsc.org}\BibitemShut {NoStop}%
\bibitem [{\citenamefont {Stankus}\ \emph {et~al.}(2020)\citenamefont {Stankus}, \citenamefont {Yong}, \citenamefont {Ruddock}, \citenamefont {Ma}, \citenamefont {Carrascosa}, \citenamefont {Goff}, \citenamefont {Boutet}, \citenamefont {Xu}, \citenamefont {Zotev}, \citenamefont {Kirrander}, \citenamefont {Minitti},\ and\ \citenamefont {Weber}}]{Stankus2020review}%
  \BibitemOpen
  \bibfield  {author} {\bibinfo {author} {\bibfnamefont {B.}~\bibnamefont {Stankus}}, \bibinfo {author} {\bibfnamefont {H.}~\bibnamefont {Yong}}, \bibinfo {author} {\bibfnamefont {J.}~\bibnamefont {Ruddock}}, \bibinfo {author} {\bibfnamefont {L.}~\bibnamefont {Ma}}, \bibinfo {author} {\bibfnamefont {A.~M.}\ \bibnamefont {Carrascosa}}, \bibinfo {author} {\bibfnamefont {N.}~\bibnamefont {Goff}}, \bibinfo {author} {\bibfnamefont {S.}~\bibnamefont {Boutet}}, \bibinfo {author} {\bibfnamefont {X.}~\bibnamefont {Xu}}, \bibinfo {author} {\bibfnamefont {N.}~\bibnamefont {Zotev}}, \bibinfo {author} {\bibfnamefont {A.}~\bibnamefont {Kirrander}}, \bibinfo {author} {\bibfnamefont {M.}~\bibnamefont {Minitti}}, \ and\ \bibinfo {author} {\bibfnamefont {P.~M.}\ \bibnamefont {Weber}},\ }\bibfield  {title} {\enquote {\bibinfo {title} {Advances in {Ultrafast} {Gas}-{Phase} {X}-ray {Scattering}},}\ }\href {\doibase 10.1088/1361-6455/abbfea} {\bibfield  {journal} {\bibinfo  {journal} {J. Phys. B}\ }\textbf {\bibinfo {volume}
  {53}},\ \bibinfo {pages} {234004} (\bibinfo {year} {2020})}\BibitemShut {NoStop}%
\bibitem [{\citenamefont {Yang}\ \emph {et~al.}(2014)\citenamefont {Yang}, \citenamefont {Makhija}, \citenamefont {Kumarappan},\ and\ \citenamefont {Centurion}}]{yang_reconstruction_2014}%
  \BibitemOpen
  \bibfield  {author} {\bibinfo {author} {\bibfnamefont {J.}~\bibnamefont {Yang}}, \bibinfo {author} {\bibfnamefont {V.}~\bibnamefont {Makhija}}, \bibinfo {author} {\bibfnamefont {V.}~\bibnamefont {Kumarappan}}, \ and\ \bibinfo {author} {\bibfnamefont {M.}~\bibnamefont {Centurion}},\ }\bibfield  {title} {\enquote {\bibinfo {title} {Reconstruction of three-dimensional molecular structure from diffraction of laser-aligned molecules},}\ }\href {\doibase 10.1063/1.4889840} {\bibfield  {journal} {\bibinfo  {journal} {Struct. Dyn.}\ }\textbf {\bibinfo {volume} {1}},\ \bibinfo {pages} {044101} (\bibinfo {year} {2014})}\BibitemShut {NoStop}%
\bibitem [{\citenamefont {Ishikawa}\ \emph {et~al.}(2015)\citenamefont {Ishikawa}, \citenamefont {Hayes}, \citenamefont {Keskin}, \citenamefont {Corthey}, \citenamefont {Hada}, \citenamefont {Pichugin}, \citenamefont {Marx}, \citenamefont {Hirscht}, \citenamefont {Shionuma}, \citenamefont {Onda}, \citenamefont {Okimoto}, \citenamefont {Koshihara}, \citenamefont {Yamamoto}, \citenamefont {Cui}, \citenamefont {Nomura}, \citenamefont {Oshima}, \citenamefont {Abdel-Jawad}, \citenamefont {Kato},\ and\ \citenamefont {Miller}}]{ishikawa_direct_2015}%
  \BibitemOpen
  \bibfield  {author} {\bibinfo {author} {\bibfnamefont {T.}~\bibnamefont {Ishikawa}}, \bibinfo {author} {\bibfnamefont {S.~A.}\ \bibnamefont {Hayes}}, \bibinfo {author} {\bibfnamefont {S.}~\bibnamefont {Keskin}}, \bibinfo {author} {\bibfnamefont {G.}~\bibnamefont {Corthey}}, \bibinfo {author} {\bibfnamefont {M.}~\bibnamefont {Hada}}, \bibinfo {author} {\bibfnamefont {K.}~\bibnamefont {Pichugin}}, \bibinfo {author} {\bibfnamefont {A.}~\bibnamefont {Marx}}, \bibinfo {author} {\bibfnamefont {J.}~\bibnamefont {Hirscht}}, \bibinfo {author} {\bibfnamefont {K.}~\bibnamefont {Shionuma}}, \bibinfo {author} {\bibfnamefont {K.}~\bibnamefont {Onda}}, \bibinfo {author} {\bibfnamefont {Y.}~\bibnamefont {Okimoto}}, \bibinfo {author} {\bibfnamefont {S.-y.}\ \bibnamefont {Koshihara}}, \bibinfo {author} {\bibfnamefont {T.}~\bibnamefont {Yamamoto}}, \bibinfo {author} {\bibfnamefont {H.}~\bibnamefont {Cui}}, \bibinfo {author} {\bibfnamefont {M.}~\bibnamefont {Nomura}}, \bibinfo {author} {\bibfnamefont {Y.}~\bibnamefont
  {Oshima}}, \bibinfo {author} {\bibfnamefont {M.}~\bibnamefont {Abdel-Jawad}}, \bibinfo {author} {\bibfnamefont {R.}~\bibnamefont {Kato}}, \ and\ \bibinfo {author} {\bibfnamefont {R.~J.~D.}\ \bibnamefont {Miller}},\ }\bibfield  {title} {\enquote {\bibinfo {title} {Direct observation of collective modes coupled to molecular orbital–driven charge transfer},}\ }\href {\doibase 10.1126/science.aab3480} {\bibfield  {journal} {\bibinfo  {journal} {Science}\ }\textbf {\bibinfo {volume} {350}},\ \bibinfo {pages} {1501--1505} (\bibinfo {year} {2015})}\BibitemShut {NoStop}%
\bibitem [{\citenamefont {Asenov}\ \emph {et~al.}(2020)\citenamefont {Asenov}, \citenamefont {Ramamoorthy}, \citenamefont {Zotev},\ and\ \citenamefont {Kirrander}}]{Asenov2020}%
  \BibitemOpen
  \bibfield  {author} {\bibinfo {author} {\bibfnamefont {M.}~\bibnamefont {Asenov}}, \bibinfo {author} {\bibfnamefont {S.}~\bibnamefont {Ramamoorthy}}, \bibinfo {author} {\bibfnamefont {N.}~\bibnamefont {Zotev}}, \ and\ \bibinfo {author} {\bibfnamefont {A.}~\bibnamefont {Kirrander}},\ }\bibfield  {title} {\enquote {\bibinfo {title} {Inversion of {Ultrafast} {X}-ray {Scattering} with {Dynamics} {Constraints}},}\ }in\ \href {https://ml4physicalsciences.github.io/2020/} {\emph {\bibinfo {booktitle} {Machine {Learning} and the {Physical} {Sciences}}}}\ (\bibinfo {year} {2020})\ p.~\bibinfo {pages} {7}\BibitemShut {NoStop}%
\bibitem [{\citenamefont {Yong}\ \emph {et~al.}(2021)\citenamefont {Yong}, \citenamefont {Carrascosa}, \citenamefont {Ma}, \citenamefont {Stankus}, \citenamefont {Minitti}, \citenamefont {Kirrander},\ and\ \citenamefont {Weber}}]{YongFD2021}%
  \BibitemOpen
  \bibfield  {author} {\bibinfo {author} {\bibfnamefont {H.}~\bibnamefont {Yong}}, \bibinfo {author} {\bibfnamefont {A.~M.}\ \bibnamefont {Carrascosa}}, \bibinfo {author} {\bibfnamefont {L.}~\bibnamefont {Ma}}, \bibinfo {author} {\bibfnamefont {B.}~\bibnamefont {Stankus}}, \bibinfo {author} {\bibfnamefont {M.~P.}\ \bibnamefont {Minitti}}, \bibinfo {author} {\bibfnamefont {A.}~\bibnamefont {Kirrander}}, \ and\ \bibinfo {author} {\bibfnamefont {P.~M.}\ \bibnamefont {Weber}},\ }\bibfield  {title} {\enquote {\bibinfo {title} {Determination of excited state molecular structures from time-resolved gas-phase {X}-ray scattering},}\ }\href {\doibase 10.1039/D0FD00118J} {\bibfield  {journal} {\bibinfo  {journal} {Faraday Disc.}\ }\textbf {\bibinfo {volume} {228}},\ \bibinfo {pages} {104--122} (\bibinfo {year} {2021})}\BibitemShut {NoStop}%
\bibitem [{\citenamefont {Acheson}\ and\ \citenamefont {Kirrander}(2023)}]{Acheson2023inversion}%
  \BibitemOpen
  \bibfield  {author} {\bibinfo {author} {\bibfnamefont {K.}~\bibnamefont {Acheson}}\ and\ \bibinfo {author} {\bibfnamefont {A.}~\bibnamefont {Kirrander}},\ }\bibfield  {title} {\enquote {\bibinfo {title} {Robust {Inversion} of {Time}-{Resolved} {Data} via {Forward}-{Optimization} in a {Trajectory} {Basis}},}\ }\href {\doibase 10.1021/acs.jctc.2c01113} {\bibfield  {journal} {\bibinfo  {journal} {J. Chem. Theory Comp.}\ }\textbf {\bibinfo {volume} {19}},\ \bibinfo {pages} {2721--2734} (\bibinfo {year} {2023})}\BibitemShut {NoStop}%
\bibitem [{\citenamefont {Drury-Lessard}\ and\ \citenamefont {Moule}(1978)}]{Drury-Lessard1978}%
  \BibitemOpen
  \bibfield  {author} {\bibinfo {author} {\bibfnamefont {C.~R.}\ \bibnamefont {Drury-Lessard}}\ and\ \bibinfo {author} {\bibfnamefont {D.~C.}\ \bibnamefont {Moule}},\ }\bibfield  {title} {\enquote {\bibinfo {title} {{Ring puckering in the 1 B 2( n , 3 s ) Rydberg electronic state of cyclobutanone}},}\ }\href {\doibase 10.1063/1.435703} {\bibfield  {journal} {\bibinfo  {journal} {The Journal of Chemical Physics}\ }\textbf {\bibinfo {volume} {68}},\ \bibinfo {pages} {5392--5395} (\bibinfo {year} {1978})}\BibitemShut {NoStop}%
\bibitem [{\citenamefont {O'Toole}\ \emph {et~al.}(1991)\citenamefont {O'Toole}, \citenamefont {Brint}, \citenamefont {Kosmidis}, \citenamefont {Boulakis},\ and\ \citenamefont {Tsekeris}}]{OToole1991}%
  \BibitemOpen
  \bibfield  {author} {\bibinfo {author} {\bibfnamefont {L.}~\bibnamefont {O'Toole}}, \bibinfo {author} {\bibfnamefont {P.}~\bibnamefont {Brint}}, \bibinfo {author} {\bibfnamefont {C.}~\bibnamefont {Kosmidis}}, \bibinfo {author} {\bibfnamefont {G.}~\bibnamefont {Boulakis}}, \ and\ \bibinfo {author} {\bibfnamefont {P.}~\bibnamefont {Tsekeris}},\ }\bibfield  {title} {\enquote {\bibinfo {title} {{Vacuum-ultraviolet absorption spectra of propanone, butanone and the cyclic ketones C n H 2n–2 O (n= 4, 5, 6, 7)}},}\ }\href {\doibase 10.1039/FT9918703343} {\bibfield  {journal} {\bibinfo  {journal} {J. Chem. Soc., Faraday Trans.}\ }\textbf {\bibinfo {volume} {87}},\ \bibinfo {pages} {3343--3351} (\bibinfo {year} {1991})}\BibitemShut {NoStop}%
\bibitem [{\citenamefont {Whitlock}\ and\ \citenamefont {Duncan}(1971)}]{Whitlock1971}%
  \BibitemOpen
  \bibfield  {author} {\bibinfo {author} {\bibfnamefont {R.~F.}\ \bibnamefont {Whitlock}}\ and\ \bibinfo {author} {\bibfnamefont {A.~B.}\ \bibnamefont {Duncan}},\ }\bibfield  {title} {\enquote {\bibinfo {title} {{Electronic spectrum of cyclobutanone}},}\ }\href {\doibase 10.1063/1.1675511} {\bibfield  {journal} {\bibinfo  {journal} {The Journal of Chemical Physics}\ }\textbf {\bibinfo {volume} {55}},\ \bibinfo {pages} {218--224} (\bibinfo {year} {1971})}\BibitemShut {NoStop}%
\bibitem [{\citenamefont {Kuhlman}, \citenamefont {S{\o}lling},\ and\ \citenamefont {M{\o}ller}(2012)}]{Kuhlman2012}%
  \BibitemOpen
  \bibfield  {author} {\bibinfo {author} {\bibfnamefont {T.~S.}\ \bibnamefont {Kuhlman}}, \bibinfo {author} {\bibfnamefont {T.~I.}\ \bibnamefont {S{\o}lling}}, \ and\ \bibinfo {author} {\bibfnamefont {K.~B.}\ \bibnamefont {M{\o}ller}},\ }\bibfield  {title} {\enquote {\bibinfo {title} {{Coherent Motion Reveals Non‐Ergodic Nature of Internal Conversion between Excited States}},}\ }\href {\doibase 10.1002/cphc.201100929} {\bibfield  {journal} {\bibinfo  {journal} {ChemPhysChem}\ }\textbf {\bibinfo {volume} {13}},\ \bibinfo {pages} {820--827} (\bibinfo {year} {2012})}\BibitemShut {NoStop}%
\bibitem [{\citenamefont {Benson}\ and\ \citenamefont {Kistiakowsky}(1942)}]{Benson1942}%
  \BibitemOpen
  \bibfield  {author} {\bibinfo {author} {\bibfnamefont {S.~W.}\ \bibnamefont {Benson}}\ and\ \bibinfo {author} {\bibfnamefont {G.~B.}\ \bibnamefont {Kistiakowsky}},\ }\bibfield  {title} {\enquote {\bibinfo {title} {{The Photochemical Decomposition of Cyclic Ketones}},}\ }\href {\doibase 10.1021/ja01253a021} {\bibfield  {journal} {\bibinfo  {journal} {Journal of the American Chemical Society}\ }\textbf {\bibinfo {volume} {64}},\ \bibinfo {pages} {80--86} (\bibinfo {year} {1942})}\BibitemShut {NoStop}%
\bibitem [{\citenamefont {Denschlag}\ and\ \citenamefont {Lee}(1968)}]{Denschlag1968}%
  \BibitemOpen
  \bibfield  {author} {\bibinfo {author} {\bibfnamefont {H.~O.}\ \bibnamefont {Denschlag}}\ and\ \bibinfo {author} {\bibfnamefont {E.~K.~C.}\ \bibnamefont {Lee}},\ }\bibfield  {title} {\enquote {\bibinfo {title} {{Benzene photosensitization and direct photolysis of cyclobutanone and cyclobutanone-2-t in the gas phase}},}\ }\href {\doibase 10.1021/ja01016a005} {\bibfield  {journal} {\bibinfo  {journal} {Journal of the American Chemical Society}\ }\textbf {\bibinfo {volume} {90}},\ \bibinfo {pages} {3628--3638} (\bibinfo {year} {1968})}\BibitemShut {NoStop}%
\bibitem [{\citenamefont {Lee}\ and\ \citenamefont {Lee}(1969)}]{Lee1969}%
  \BibitemOpen
  \bibfield  {author} {\bibinfo {author} {\bibfnamefont {N.~E.}\ \bibnamefont {Lee}}\ and\ \bibinfo {author} {\bibfnamefont {E.~K.~C.}\ \bibnamefont {Lee}},\ }\bibfield  {title} {\enquote {\bibinfo {title} {{Tracer Study of Photochemically Excited Cyclobutanone-2- t and Cyclobutanone. II. Detailed Mechanism, Energetics, Unimolecular Decomposition Rates, and Intermolecular Vibrational Energy Transfer}},}\ }\href {\doibase 10.1063/1.1671339} {\bibfield  {journal} {\bibinfo  {journal} {The Journal of Chemical Physics}\ }\textbf {\bibinfo {volume} {50}},\ \bibinfo {pages} {2094--2107} (\bibinfo {year} {1969})}\BibitemShut {NoStop}%
\bibitem [{\citenamefont {Lee}, \citenamefont {Shortridge},\ and\ \citenamefont {Rusbult}(1971)}]{Lee1971}%
  \BibitemOpen
  \bibfield  {author} {\bibinfo {author} {\bibfnamefont {E.~K.~C.}\ \bibnamefont {Lee}}, \bibinfo {author} {\bibfnamefont {R.~G.}\ \bibnamefont {Shortridge}}, \ and\ \bibinfo {author} {\bibfnamefont {C.~F.}\ \bibnamefont {Rusbult}},\ }\bibfield  {title} {\enquote {\bibinfo {title} {{Fluorescence excitation study of cyclobutanone, cyclopentanone, and cyclohexanone in the gas phase}},}\ }\href {\doibase 10.1021/ja00737a005} {\bibfield  {journal} {\bibinfo  {journal} {Journal of the American Chemical Society}\ }\textbf {\bibinfo {volume} {93}},\ \bibinfo {pages} {1863--1867} (\bibinfo {year} {1971})}\BibitemShut {NoStop}%
\bibitem [{\citenamefont {Diau}, \citenamefont {K{\"{o}}tting},\ and\ \citenamefont {Zewail}(2001)}]{Diau2001}%
  \BibitemOpen
  \bibfield  {author} {\bibinfo {author} {\bibfnamefont {E.~W.}\ \bibnamefont {Diau}}, \bibinfo {author} {\bibfnamefont {G.}~\bibnamefont {K{\"{o}}tting}}, \ and\ \bibinfo {author} {\bibfnamefont {A.~H.}\ \bibnamefont {Zewail}},\ }\bibfield  {title} {\enquote {\bibinfo {title} {{Femtochemistry of norrish type-1 reactions: ii. the anomalous predissociation dynamics of cyclobutanone on the S1 surface}},}\ }\href {\doibase 10.1002/1439-7641(20010518)2:5<294::aid-cphc294>3.0.co;2-5} {\bibfield  {journal} {\bibinfo  {journal} {ChemPhysChem}\ }\textbf {\bibinfo {volume} {2}},\ \bibinfo {pages} {294--309} (\bibinfo {year} {2001})}\BibitemShut {NoStop}%
\bibitem [{\citenamefont {Kao}\ \emph {et~al.}(2020)\citenamefont {Kao}, \citenamefont {Venkatraman}, \citenamefont {Ashfold},\ and\ \citenamefont {Orr-Ewing}}]{Kao2020}%
  \BibitemOpen
  \bibfield  {author} {\bibinfo {author} {\bibfnamefont {M.~H.}\ \bibnamefont {Kao}}, \bibinfo {author} {\bibfnamefont {R.~K.}\ \bibnamefont {Venkatraman}}, \bibinfo {author} {\bibfnamefont {M.~N.}\ \bibnamefont {Ashfold}}, \ and\ \bibinfo {author} {\bibfnamefont {A.~J.}\ \bibnamefont {Orr-Ewing}},\ }\bibfield  {title} {\enquote {\bibinfo {title} {{Effects of ring-strain on the ultrafast photochemistry of cyclic ketones}},}\ }\href {\doibase 10.1039/c9sc05208a} {\bibfield  {journal} {\bibinfo  {journal} {Chemical Science}\ }\textbf {\bibinfo {volume} {11}},\ \bibinfo {pages} {1991--2000} (\bibinfo {year} {2020})}\BibitemShut {NoStop}%
\bibitem [{\citenamefont {Makhov}\ \emph {et~al.}(2014)\citenamefont {Makhov}, \citenamefont {Glover}, \citenamefont {Martinez},\ and\ \citenamefont {Shalashilin}}]{AIMC}%
  \BibitemOpen
  \bibfield  {author} {\bibinfo {author} {\bibfnamefont {D.~V.}\ \bibnamefont {Makhov}}, \bibinfo {author} {\bibfnamefont {W.~J.}\ \bibnamefont {Glover}}, \bibinfo {author} {\bibfnamefont {T.~J.}\ \bibnamefont {Martinez}}, \ and\ \bibinfo {author} {\bibfnamefont {D.~V.}\ \bibnamefont {Shalashilin}},\ }\bibfield  {title} {\enquote {\bibinfo {title} {{Ab initio multiple cloning algorithm for quantum nonadiabatic molecular dynamics}},}\ }\href {\doibase 10.1063/1.4891530} {\bibfield  {journal} {\bibinfo  {journal} {The Journal of Chemical Physics}\ }\textbf {\bibinfo {volume} {141}},\ \bibinfo {pages} {054110} (\bibinfo {year} {2014})},\ \Eprint {http://arxiv.org/abs/https://pubs.aip.org/aip/jcp/article-pdf/doi/10.1063/1.4891530/13375170/054110\_1\_online.pdf} {https://pubs.aip.org/aip/jcp/article-pdf/doi/10.1063/1.4891530/13375170/054110\_1\_online.pdf} \BibitemShut {NoStop}%
\bibitem [{\citenamefont {Makhov}\ \emph {et~al.}(2017)\citenamefont {Makhov}, \citenamefont {Symonds}, \citenamefont {Fernandez-Alberti},\ and\ \citenamefont {Shalashilin}}]{MAKHOV2017200}%
  \BibitemOpen
  \bibfield  {author} {\bibinfo {author} {\bibfnamefont {D.~V.}\ \bibnamefont {Makhov}}, \bibinfo {author} {\bibfnamefont {C.}~\bibnamefont {Symonds}}, \bibinfo {author} {\bibfnamefont {S.}~\bibnamefont {Fernandez-Alberti}}, \ and\ \bibinfo {author} {\bibfnamefont {D.~V.}\ \bibnamefont {Shalashilin}},\ }\bibfield  {title} {\enquote {\bibinfo {title} {Ab initio quantum direct dynamics simulations of ultrafast photochemistry with multiconfigurational ehrenfest approach},}\ }\href {\doibase https://doi.org/10.1016/j.chemphys.2017.04.003} {\bibfield  {journal} {\bibinfo  {journal} {Chemical Physics}\ }\textbf {\bibinfo {volume} {493}},\ \bibinfo {pages} {200--218} (\bibinfo {year} {2017})}\BibitemShut {NoStop}%
\bibitem [{\citenamefont {Freixas}\ \emph {et~al.}(2018)\citenamefont {Freixas}, \citenamefont {Fernandez-Alberti}, \citenamefont {Makhov}, \citenamefont {Tretiak},\ and\ \citenamefont {Shalashilin}}]{C8CP02321B}%
  \BibitemOpen
  \bibfield  {author} {\bibinfo {author} {\bibfnamefont {V.~M.}\ \bibnamefont {Freixas}}, \bibinfo {author} {\bibfnamefont {S.}~\bibnamefont {Fernandez-Alberti}}, \bibinfo {author} {\bibfnamefont {D.~V.}\ \bibnamefont {Makhov}}, \bibinfo {author} {\bibfnamefont {S.}~\bibnamefont {Tretiak}}, \ and\ \bibinfo {author} {\bibfnamefont {D.}~\bibnamefont {Shalashilin}},\ }\bibfield  {title} {\enquote {\bibinfo {title} {An ab initio multiple cloning approach for the simulation of photoinduced dynamics in conjugated molecules},}\ }\href {\doibase 10.1039/C8CP02321B} {\bibfield  {journal} {\bibinfo  {journal} {Phys. Chem. Chem. Phys.}\ }\textbf {\bibinfo {volume} {20}},\ \bibinfo {pages} {17762--17772} (\bibinfo {year} {2018})}\BibitemShut {NoStop}%
\bibitem [{\citenamefont {Makhov}\ \emph {et~al.}(2015)\citenamefont {Makhov}, \citenamefont {Saita}, \citenamefont {Martinez},\ and\ \citenamefont {Shalashilin}}]{Pyrrole}%
  \BibitemOpen
  \bibfield  {author} {\bibinfo {author} {\bibfnamefont {D.~V.}\ \bibnamefont {Makhov}}, \bibinfo {author} {\bibfnamefont {K.}~\bibnamefont {Saita}}, \bibinfo {author} {\bibfnamefont {T.~J.}\ \bibnamefont {Martinez}}, \ and\ \bibinfo {author} {\bibfnamefont {D.~V.}\ \bibnamefont {Shalashilin}},\ }\bibfield  {title} {\enquote {\bibinfo {title} {Ab initio multiple cloning simulations of pyrrole photodissociation: Tker spectra and velocity map imaging},}\ }\href {\doibase 10.1039/C4CP04571H} {\bibfield  {journal} {\bibinfo  {journal} {Phys. Chem. Chem. Phys.}\ }\textbf {\bibinfo {volume} {17}},\ \bibinfo {pages} {3316--3325} (\bibinfo {year} {2015})}\BibitemShut {NoStop}%
\bibitem [{\citenamefont {Green}\ \emph {et~al.}(2019)\citenamefont {Green}, \citenamefont {Makhov}, \citenamefont {Cole-Filipiak}, \citenamefont {Symonds}, \citenamefont {Stavros},\ and\ \citenamefont {Shalashilin}}]{2EP}%
  \BibitemOpen
  \bibfield  {author} {\bibinfo {author} {\bibfnamefont {J.~A.}\ \bibnamefont {Green}}, \bibinfo {author} {\bibfnamefont {D.~V.}\ \bibnamefont {Makhov}}, \bibinfo {author} {\bibfnamefont {N.~C.}\ \bibnamefont {Cole-Filipiak}}, \bibinfo {author} {\bibfnamefont {C.}~\bibnamefont {Symonds}}, \bibinfo {author} {\bibfnamefont {V.~G.}\ \bibnamefont {Stavros}}, \ and\ \bibinfo {author} {\bibfnamefont {D.~V.}\ \bibnamefont {Shalashilin}},\ }\bibfield  {title} {\enquote {\bibinfo {title} {Ultrafast photodissociation dynamics of 2-ethylpyrrole: adding insight to experiment with ab initio multiple cloning},}\ }\href {\doibase 10.1039/C8CP06359A} {\bibfield  {journal} {\bibinfo  {journal} {Phys. Chem. Chem. Phys.}\ }\textbf {\bibinfo {volume} {21}},\ \bibinfo {pages} {3832--3841} (\bibinfo {year} {2019})}\BibitemShut {NoStop}%
\bibitem [{\citenamefont {Symonds}\ \emph {et~al.}(2019)\citenamefont {Symonds}, \citenamefont {Makhov}, \citenamefont {Cole-Filipiak}, \citenamefont {Green}, \citenamefont {Stavros},\ and\ \citenamefont {Shalashilin}}]{pyrazole}%
  \BibitemOpen
  \bibfield  {author} {\bibinfo {author} {\bibfnamefont {C.~C.}\ \bibnamefont {Symonds}}, \bibinfo {author} {\bibfnamefont {D.~V.}\ \bibnamefont {Makhov}}, \bibinfo {author} {\bibfnamefont {N.~C.}\ \bibnamefont {Cole-Filipiak}}, \bibinfo {author} {\bibfnamefont {J.~A.}\ \bibnamefont {Green}}, \bibinfo {author} {\bibfnamefont {V.~G.}\ \bibnamefont {Stavros}}, \ and\ \bibinfo {author} {\bibfnamefont {D.~V.}\ \bibnamefont {Shalashilin}},\ }\bibfield  {title} {\enquote {\bibinfo {title} {Ultrafast photodissociation dynamics of pyrazole{,} imidazole and their deuterated derivatives using ab initio multiple cloning},}\ }\href {\doibase 10.1039/C9CP00039A} {\bibfield  {journal} {\bibinfo  {journal} {Phys. Chem. Chem. Phys.}\ }\textbf {\bibinfo {volume} {21}},\ \bibinfo {pages} {9987--9995} (\bibinfo {year} {2019})}\BibitemShut {NoStop}%
\bibitem [{\citenamefont {Makhov}\ and\ \citenamefont {Shalashilin}(2021)}]{Vib}%
  \BibitemOpen
  \bibfield  {author} {\bibinfo {author} {\bibfnamefont {D.~V.}\ \bibnamefont {Makhov}}\ and\ \bibinfo {author} {\bibfnamefont {D.~V.}\ \bibnamefont {Shalashilin}},\ }\bibfield  {title} {\enquote {\bibinfo {title} {{Simulation of the effect of vibrational pre-excitation on the dynamics of pyrrole photo-dissociation}},}\ }\href {\doibase 10.1063/5.0040178} {\bibfield  {journal} {\bibinfo  {journal} {The Journal of Chemical Physics}\ }\textbf {\bibinfo {volume} {154}},\ \bibinfo {pages} {104119} (\bibinfo {year} {2021})},\ \Eprint {http://arxiv.org/abs/https://pubs.aip.org/aip/jcp/article-pdf/doi/10.1063/5.0040178/13469571/104119\_1\_online.pdf} {https://pubs.aip.org/aip/jcp/article-pdf/doi/10.1063/5.0040178/13469571/104119\_1\_online.pdf} \BibitemShut {NoStop}%
\bibitem [{\citenamefont {Makhov}\ \emph {et~al.}(2022)\citenamefont {Makhov}, \citenamefont {Adeyemi}, \citenamefont {Cowperthwaite},\ and\ \citenamefont {Shalashilin}}]{Vib_D}%
  \BibitemOpen
  \bibfield  {author} {\bibinfo {author} {\bibfnamefont {D.~V.}\ \bibnamefont {Makhov}}, \bibinfo {author} {\bibfnamefont {S.}~\bibnamefont {Adeyemi}}, \bibinfo {author} {\bibfnamefont {M.}~\bibnamefont {Cowperthwaite}}, \ and\ \bibinfo {author} {\bibfnamefont {D.~V.}\ \bibnamefont {Shalashilin}},\ }\bibfield  {title} {\enquote {\bibinfo {title} {Simulation of the dynamics of vibrationally mediated photodissociation for deuterated pyrrole},}\ }\href {\doibase 10.1088/2399-6528/ac4d39} {\bibfield  {journal} {\bibinfo  {journal} {Journal of Physics Communications}\ }\textbf {\bibinfo {volume} {6}},\ \bibinfo {pages} {025001} (\bibinfo {year} {2022})}\BibitemShut {NoStop}%
\bibitem [{\citenamefont {Shalashilin}(2009)}]{DS2009}%
  \BibitemOpen
  \bibfield  {author} {\bibinfo {author} {\bibfnamefont {D.~V.}\ \bibnamefont {Shalashilin}},\ }\bibfield  {title} {\enquote {\bibinfo {title} {{Quantum mechanics with the basis set guided by Ehrenfest trajectories: Theory and application to spin-boson model}},}\ }\href {\doibase 10.1063/1.3153302} {\bibfield  {journal} {\bibinfo  {journal} {The Journal of Chemical Physics}\ }\textbf {\bibinfo {volume} {130}},\ \bibinfo {pages} {244101} (\bibinfo {year} {2009})},\ \Eprint {http://arxiv.org/abs/https://pubs.aip.org/aip/jcp/article-pdf/doi/10.1063/1.3153302/15995187/244101\_1\_online.pdf} {https://pubs.aip.org/aip/jcp/article-pdf/doi/10.1063/1.3153302/15995187/244101\_1\_online.pdf} \BibitemShut {NoStop}%
\bibitem [{\citenamefont {Shalashilin}(2010)}]{DS2010}%
  \BibitemOpen
  \bibfield  {author} {\bibinfo {author} {\bibfnamefont {D.~V.}\ \bibnamefont {Shalashilin}},\ }\bibfield  {title} {\enquote {\bibinfo {title} {{Nonadiabatic dynamics with the help of multiconfigurational Ehrenfest method: Improved theory and fully quantum 24D simulation of pyrazine}},}\ }\href {\doibase 10.1063/1.3442747} {\bibfield  {journal} {\bibinfo  {journal} {The Journal of Chemical Physics}\ }\textbf {\bibinfo {volume} {132}},\ \bibinfo {pages} {244111} (\bibinfo {year} {2010})},\ \Eprint {http://arxiv.org/abs/https://pubs.aip.org/aip/jcp/article-pdf/doi/10.1063/1.3442747/16122321/244111\_1\_online.pdf} {https://pubs.aip.org/aip/jcp/article-pdf/doi/10.1063/1.3442747/16122321/244111\_1\_online.pdf} \BibitemShut {NoStop}%
\bibitem [{\citenamefont {Shalashilin}(2011)}]{DS2011}%
  \BibitemOpen
  \bibfield  {author} {\bibinfo {author} {\bibfnamefont {D.~V.}\ \bibnamefont {Shalashilin}},\ }\bibfield  {title} {\enquote {\bibinfo {title} {Multiconfigurational ehrenfest approach to quantum coherent dynamics in large molecular systems},}\ }\href {\doibase 10.1039/C1FD00034A} {\bibfield  {journal} {\bibinfo  {journal} {Faraday Discuss.}\ }\textbf {\bibinfo {volume} {153}},\ \bibinfo {pages} {105--116} (\bibinfo {year} {2011})}\BibitemShut {NoStop}%
\bibitem [{\citenamefont {Saita}\ and\ \citenamefont {Shalashilin}(2012)}]{Saita}%
  \BibitemOpen
  \bibfield  {author} {\bibinfo {author} {\bibfnamefont {K.}~\bibnamefont {Saita}}\ and\ \bibinfo {author} {\bibfnamefont {D.~V.}\ \bibnamefont {Shalashilin}},\ }\bibfield  {title} {\enquote {\bibinfo {title} {{On-the-fly ab initio molecular dynamics with multiconfigurational Ehrenfest method}},}\ }\href {\doibase 10.1063/1.4734313} {\bibfield  {journal} {\bibinfo  {journal} {The Journal of Chemical Physics}\ }\textbf {\bibinfo {volume} {137}},\ \bibinfo {pages} {22A506} (\bibinfo {year} {2012})},\ \Eprint {http://arxiv.org/abs/https://pubs.aip.org/aip/jcp/article-pdf/doi/10.1063/1.4734313/14003357/22a506\_1\_online.pdf} {https://pubs.aip.org/aip/jcp/article-pdf/doi/10.1063/1.4734313/14003357/22a506\_1\_online.pdf} \BibitemShut {NoStop}%
\bibitem [{\citenamefont {Curchod}\ and\ \citenamefont {Mart{\'{i}}nez}(2018)}]{Curchod2018d}%
  \BibitemOpen
  \bibfield  {author} {\bibinfo {author} {\bibfnamefont {B.~F.~E.}\ \bibnamefont {Curchod}}\ and\ \bibinfo {author} {\bibfnamefont {T.~J.}\ \bibnamefont {Mart{\'{i}}nez}},\ }\bibfield  {title} {\enquote {\bibinfo {title} {{Ab Initio Nonadiabatic Quantum Molecular Dynamics}},}\ }\href {\doibase 10.1021/acs.chemrev.7b00423} {\bibfield  {journal} {\bibinfo  {journal} {Chemical Reviews}\ }\textbf {\bibinfo {volume} {118}},\ \bibinfo {pages} {3305--3336} (\bibinfo {year} {2018})}\BibitemShut {NoStop}%
\bibitem [{\citenamefont {Shalashilin}\ and\ \citenamefont {Child}(2008)}]{DS2008}%
  \BibitemOpen
  \bibfield  {author} {\bibinfo {author} {\bibfnamefont {D.~V.}\ \bibnamefont {Shalashilin}}\ and\ \bibinfo {author} {\bibfnamefont {M.~S.}\ \bibnamefont {Child}},\ }\bibfield  {title} {\enquote {\bibinfo {title} {{Basis set sampling in the method of coupled coherent states: Coherent state swarms, trains, and pancakes}},}\ }\href {\doibase 10.1063/1.2828509} {\bibfield  {journal} {\bibinfo  {journal} {The Journal of Chemical Physics}\ }\textbf {\bibinfo {volume} {128}},\ \bibinfo {pages} {054102} (\bibinfo {year} {2008})},\ \Eprint {http://arxiv.org/abs/https://pubs.aip.org/aip/jcp/article-pdf/doi/10.1063/1.2828509/15411378/054102\_1\_online.pdf} {https://pubs.aip.org/aip/jcp/article-pdf/doi/10.1063/1.2828509/15411378/054102\_1\_online.pdf} \BibitemShut {NoStop}%
\bibitem [{\citenamefont {Symonds}, \citenamefont {Kattirtzi},\ and\ \citenamefont {Shalashilin}(2018)}]{EFFCTSMPLNG}%
  \BibitemOpen
  \bibfield  {author} {\bibinfo {author} {\bibfnamefont {C.}~\bibnamefont {Symonds}}, \bibinfo {author} {\bibfnamefont {J.~A.}\ \bibnamefont {Kattirtzi}}, \ and\ \bibinfo {author} {\bibfnamefont {D.~V.}\ \bibnamefont {Shalashilin}},\ }\bibfield  {title} {\enquote {\bibinfo {title} {{The effect of sampling techniques used in the multiconfigurational Ehrenfest method}},}\ }\href {\doibase 10.1063/1.5020567} {\bibfield  {journal} {\bibinfo  {journal} {The Journal of Chemical Physics}\ }\textbf {\bibinfo {volume} {148}},\ \bibinfo {pages} {184113} (\bibinfo {year} {2018})},\ \Eprint {http://arxiv.org/abs/https://pubs.aip.org/aip/jcp/article-pdf/doi/10.1063/1.5020567/15541045/184113\_1\_online.pdf} {https://pubs.aip.org/aip/jcp/article-pdf/doi/10.1063/1.5020567/15541045/184113\_1\_online.pdf} \BibitemShut {NoStop}%
\bibitem [{\citenamefont {Freixas}\ \emph {et~al.}(2021)\citenamefont {Freixas}, \citenamefont {White}, \citenamefont {Nelson}, \citenamefont {Song}, \citenamefont {Makhov}, \citenamefont {Shalashilin}, \citenamefont {Fernandez-Alberti},\ and\ \citenamefont {Tretiak}}]{CmprsnCnvrg}%
  \BibitemOpen
  \bibfield  {author} {\bibinfo {author} {\bibfnamefont {V.~M.}\ \bibnamefont {Freixas}}, \bibinfo {author} {\bibfnamefont {A.~J.}\ \bibnamefont {White}}, \bibinfo {author} {\bibfnamefont {T.}~\bibnamefont {Nelson}}, \bibinfo {author} {\bibfnamefont {H.}~\bibnamefont {Song}}, \bibinfo {author} {\bibfnamefont {D.~V.}\ \bibnamefont {Makhov}}, \bibinfo {author} {\bibfnamefont {D.}~\bibnamefont {Shalashilin}}, \bibinfo {author} {\bibfnamefont {S.}~\bibnamefont {Fernandez-Alberti}}, \ and\ \bibinfo {author} {\bibfnamefont {S.}~\bibnamefont {Tretiak}},\ }\bibfield  {title} {\enquote {\bibinfo {title} {Nonadiabatic excited-state molecular dynamics methodologies: Comparison and convergence},}\ }\href {\doibase 10.1021/acs.jpclett.1c00266} {\bibfield  {journal} {\bibinfo  {journal} {The Journal of Physical Chemistry Letters}\ }\textbf {\bibinfo {volume} {12}},\ \bibinfo {pages} {2970--2982} (\bibinfo {year} {2021})},\ \bibinfo {note} {pMID: 33730495},\ \Eprint
  {http://arxiv.org/abs/https://doi.org/10.1021/acs.jpclett.1c00266} {https://doi.org/10.1021/acs.jpclett.1c00266} \BibitemShut {NoStop}%
\bibitem [{\citenamefont {Freixas}\ \emph {et~al.}(2023)\citenamefont {Freixas}, \citenamefont {Malone}, \citenamefont {Li}, \citenamefont {Song}, \citenamefont {Negrin-Yuvero}, \citenamefont {Pérez-Castillo}, \citenamefont {White}, \citenamefont {Gibson}, \citenamefont {Makhov}, \citenamefont {Shalashilin}, \citenamefont {Zhang}, \citenamefont {Fedik}, \citenamefont {Kulichenko}, \citenamefont {Messerly}, \citenamefont {Mohanam}, \citenamefont {Sharifzadeh}, \citenamefont {Bastida}, \citenamefont {Mukamel}, \citenamefont {Fernandez-Alberti},\ and\ \citenamefont {Tretiak}}]{NEXMD}%
  \BibitemOpen
  \bibfield  {author} {\bibinfo {author} {\bibfnamefont {V.~M.}\ \bibnamefont {Freixas}}, \bibinfo {author} {\bibfnamefont {W.}~\bibnamefont {Malone}}, \bibinfo {author} {\bibfnamefont {X.}~\bibnamefont {Li}}, \bibinfo {author} {\bibfnamefont {H.}~\bibnamefont {Song}}, \bibinfo {author} {\bibfnamefont {H.}~\bibnamefont {Negrin-Yuvero}}, \bibinfo {author} {\bibfnamefont {R.}~\bibnamefont {Pérez-Castillo}}, \bibinfo {author} {\bibfnamefont {A.}~\bibnamefont {White}}, \bibinfo {author} {\bibfnamefont {T.~R.}\ \bibnamefont {Gibson}}, \bibinfo {author} {\bibfnamefont {D.~V.}\ \bibnamefont {Makhov}}, \bibinfo {author} {\bibfnamefont {D.~V.}\ \bibnamefont {Shalashilin}}, \bibinfo {author} {\bibfnamefont {Y.}~\bibnamefont {Zhang}}, \bibinfo {author} {\bibfnamefont {N.}~\bibnamefont {Fedik}}, \bibinfo {author} {\bibfnamefont {M.}~\bibnamefont {Kulichenko}}, \bibinfo {author} {\bibfnamefont {R.}~\bibnamefont {Messerly}}, \bibinfo {author} {\bibfnamefont {L.~N.}\ \bibnamefont {Mohanam}}, \bibinfo {author}
  {\bibfnamefont {S.}~\bibnamefont {Sharifzadeh}}, \bibinfo {author} {\bibfnamefont {A.}~\bibnamefont {Bastida}}, \bibinfo {author} {\bibfnamefont {S.}~\bibnamefont {Mukamel}}, \bibinfo {author} {\bibfnamefont {S.}~\bibnamefont {Fernandez-Alberti}}, \ and\ \bibinfo {author} {\bibfnamefont {S.}~\bibnamefont {Tretiak}},\ }\bibfield  {title} {\enquote {\bibinfo {title} {Nexmd v2.0 software package for nonadiabatic excited state molecular dynamics simulations},}\ }\href {\doibase 10.1021/acs.jctc.3c00583} {\bibfield  {journal} {\bibinfo  {journal} {Journal of Chemical Theory and Computation}\ }\textbf {\bibinfo {volume} {19}},\ \bibinfo {pages} {5356--5368} (\bibinfo {year} {2023})},\ \bibinfo {note} {pMID: 37506288},\ \Eprint {http://arxiv.org/abs/https://doi.org/10.1021/acs.jctc.3c00583} {https://doi.org/10.1021/acs.jctc.3c00583} \BibitemShut {NoStop}%
\bibitem [{\citenamefont {Makhov}\ and\ \citenamefont {Shalashilin}(2018)}]{MAKHOV201846}%
  \BibitemOpen
  \bibfield  {author} {\bibinfo {author} {\bibfnamefont {D.~V.}\ \bibnamefont {Makhov}}\ and\ \bibinfo {author} {\bibfnamefont {D.~V.}\ \bibnamefont {Shalashilin}},\ }\bibfield  {title} {\enquote {\bibinfo {title} {Floquet hamiltonian for incorporating electronic excitation by a laser pulse into simulations of non-adiabatic dynamics},}\ }\href {\doibase https://doi.org/10.1016/j.chemphys.2018.07.048} {\bibfield  {journal} {\bibinfo  {journal} {Chemical Physics}\ }\textbf {\bibinfo {volume} {515}},\ \bibinfo {pages} {46--51} (\bibinfo {year} {2018})},\ \bibinfo {note} {ultrafast Photoinduced Processes in Polyatomic Molecules:Electronic Structure, Dynamics and Spectroscopy (Dedicated to Wolfgang Domcke on the occasion of his 70th birthday)}\BibitemShut {NoStop}%
\bibitem [{\citenamefont {Moreno~Carrascosa}\ \emph {et~al.}(2019)\citenamefont {Moreno~Carrascosa}, \citenamefont {Yong}, \citenamefont {Crittenden}, \citenamefont {Weber},\ and\ \citenamefont {Kirrander}}]{moreno_carrascosa_ab_2019}%
  \BibitemOpen
  \bibfield  {author} {\bibinfo {author} {\bibfnamefont {A.}~\bibnamefont {Moreno~Carrascosa}}, \bibinfo {author} {\bibfnamefont {H.}~\bibnamefont {Yong}}, \bibinfo {author} {\bibfnamefont {D.~L.}\ \bibnamefont {Crittenden}}, \bibinfo {author} {\bibfnamefont {P.~M.}\ \bibnamefont {Weber}}, \ and\ \bibinfo {author} {\bibfnamefont {A.}~\bibnamefont {Kirrander}},\ }\bibfield  {title} {\enquote {\bibinfo {title} {Ab {Initio} {Calculation} of {Total} {X}-ray {Scattering} from {Molecules}},}\ }\href {\doibase 10.1021/acs.jctc.9b00056} {\bibfield  {journal} {\bibinfo  {journal} {J. Chem. Theory Comput.}\ }\textbf {\bibinfo {volume} {15}},\ \bibinfo {pages} {2836--2846} (\bibinfo {year} {2019})}\BibitemShut {NoStop}%
\bibitem [{\citenamefont {Parrish}\ and\ \citenamefont {Mart{\'{i}}nez}(2019)}]{Parrish2019}%
  \BibitemOpen
  \bibfield  {author} {\bibinfo {author} {\bibfnamefont {R.~M.}\ \bibnamefont {Parrish}}\ and\ \bibinfo {author} {\bibfnamefont {T.~J.}\ \bibnamefont {Mart{\'{i}}nez}},\ }\bibfield  {title} {\enquote {\bibinfo {title} {Ab initio computation of rotationally-averaged pump-probe x-ray and electron diffraction signals},}\ }\href {\doibase 10.1021/acs.jctc.8b01051} {\bibfield  {journal} {\bibinfo  {journal} {J. Chem. Theory Comp.}\ }\textbf {\bibinfo {volume} {0}},\ \bibinfo {pages} {null} (\bibinfo {year} {2019})},\ \bibinfo {note} {pMID: 30702882},\ \Eprint {http://arxiv.org/abs/https://doi.org/10.1021/acs.jctc.8b01051} {https://doi.org/10.1021/acs.jctc.8b01051} \BibitemShut {NoStop}%
\bibitem [{\citenamefont {Zotev}\ \emph {et~al.}(2020)\citenamefont {Zotev}, \citenamefont {Moreno~Carrascosa}, \citenamefont {Simmermacher},\ and\ \citenamefont {Kirrander}}]{zotev_excited_2020}%
  \BibitemOpen
  \bibfield  {author} {\bibinfo {author} {\bibfnamefont {N.}~\bibnamefont {Zotev}}, \bibinfo {author} {\bibfnamefont {A.}~\bibnamefont {Moreno~Carrascosa}}, \bibinfo {author} {\bibfnamefont {M.}~\bibnamefont {Simmermacher}}, \ and\ \bibinfo {author} {\bibfnamefont {A.}~\bibnamefont {Kirrander}},\ }\bibfield  {title} {\enquote {\bibinfo {title} {Excited {Electronic} {States} in {Total} {Isotropic} {Scattering} from {Molecules}},}\ }\href {\doibase 10.1021/acs.jctc.9b00670} {\bibfield  {journal} {\bibinfo  {journal} {J. Chem. Theory Comput.}\ }\textbf {\bibinfo {volume} {16}},\ \bibinfo {pages} {2594--2605} (\bibinfo {year} {2020})}\BibitemShut {NoStop}%
\bibitem [{\citenamefont {Carrascosa}\ \emph {et~al.}(2022)\citenamefont {Carrascosa}, \citenamefont {Coe}, \citenamefont {Simmermacher}, \citenamefont {Paterson},\ and\ \citenamefont {Kirrander}}]{Moreno2022}%
  \BibitemOpen
  \bibfield  {author} {\bibinfo {author} {\bibfnamefont {A.~M.}\ \bibnamefont {Carrascosa}}, \bibinfo {author} {\bibfnamefont {J.~P.}\ \bibnamefont {Coe}}, \bibinfo {author} {\bibfnamefont {M.}~\bibnamefont {Simmermacher}}, \bibinfo {author} {\bibfnamefont {M.~J.}\ \bibnamefont {Paterson}}, \ and\ \bibinfo {author} {\bibfnamefont {A.}~\bibnamefont {Kirrander}},\ }\bibfield  {title} {\enquote {\bibinfo {title} {Towards high-resolution {X}-ray scattering as a probe of electron correlation},}\ }\href {\doibase 10.1039/D2CP02933B} {\bibfield  {journal} {\bibinfo  {journal} {Phys. Chem. Chem. Phys.}\ }\textbf {\bibinfo {volume} {24}},\ \bibinfo {pages} {24542--24552} (\bibinfo {year} {2022})}\BibitemShut {NoStop}%
\bibitem [{\citenamefont {Kirrander}, \citenamefont {Saita},\ and\ \citenamefont {Shalashilin}(2016)}]{kirrander_ultrafast_2016}%
  \BibitemOpen
  \bibfield  {author} {\bibinfo {author} {\bibfnamefont {A.}~\bibnamefont {Kirrander}}, \bibinfo {author} {\bibfnamefont {K.}~\bibnamefont {Saita}}, \ and\ \bibinfo {author} {\bibfnamefont {D.~V.}\ \bibnamefont {Shalashilin}},\ }\bibfield  {title} {\enquote {\bibinfo {title} {Ultrafast {X}-ray {Scattering} from {Molecules}},}\ }\href {\doibase 10.1021/acs.jctc.5b01042} {\bibfield  {journal} {\bibinfo  {journal} {J. Chem. Theory Comput.}\ }\textbf {\bibinfo {volume} {12}},\ \bibinfo {pages} {957--967} (\bibinfo {year} {2016})}\BibitemShut {NoStop}%
\bibitem [{\citenamefont {Stefanou}\ \emph {et~al.}(2017)\citenamefont {Stefanou}, \citenamefont {Saita}, \citenamefont {Shalashilin},\ and\ \citenamefont {Kirrander}}]{Minas2017}%
  \BibitemOpen
  \bibfield  {author} {\bibinfo {author} {\bibfnamefont {M.}~\bibnamefont {Stefanou}}, \bibinfo {author} {\bibfnamefont {K.}~\bibnamefont {Saita}}, \bibinfo {author} {\bibfnamefont {D.~V.}\ \bibnamefont {Shalashilin}}, \ and\ \bibinfo {author} {\bibfnamefont {A.}~\bibnamefont {Kirrander}},\ }\bibfield  {title} {\enquote {\bibinfo {title} {Comparison of ultrafast electron and x-ray diffraction – a computational study},}\ }\href {\doibase 10.1016/j.cplett.2017.03.007} {\bibfield  {journal} {\bibinfo  {journal} {Chem. Phys. Lett.}\ }\textbf {\bibinfo {volume} {683}},\ \bibinfo {pages} {300--305} (\bibinfo {year} {2017})}\BibitemShut {NoStop}%
\bibitem [{\citenamefont {Kirrander}\ and\ \citenamefont {Weber}(2017)}]{kirrander_fundamental_2017}%
  \BibitemOpen
  \bibfield  {author} {\bibinfo {author} {\bibfnamefont {A.}~\bibnamefont {Kirrander}}\ and\ \bibinfo {author} {\bibfnamefont {P.~M.}\ \bibnamefont {Weber}},\ }\bibfield  {title} {\enquote {\bibinfo {title} {Fundamental {Limits} on {Spatial} {Resolution} in {Ultrafast} {X}-ray {Diffraction}},}\ }\href {\doibase 10.3390/app7060534} {\bibfield  {journal} {\bibinfo  {journal} {Appl. Sci.}\ }\textbf {\bibinfo {volume} {7}},\ \bibinfo {pages} {534} (\bibinfo {year} {2017})}\BibitemShut {NoStop}%
\bibitem [{\citenamefont {Mott}\ and\ \citenamefont {Bragg}(1930)}]{mott_scattering_1930}%
  \BibitemOpen
  \bibfield  {author} {\bibinfo {author} {\bibfnamefont {N.~F.}\ \bibnamefont {Mott}}\ and\ \bibinfo {author} {\bibfnamefont {W.~L.}\ \bibnamefont {Bragg}},\ }\bibfield  {title} {\enquote {\bibinfo {title} {The scattering of electrons by atoms},}\ }\href {\doibase 10.1098/rspa.1930.0082} {\bibfield  {journal} {\bibinfo  {journal} {Proc. R. Soc., Lond., Ser. A}\ }\textbf {\bibinfo {volume} {127}},\ \bibinfo {pages} {658--665} (\bibinfo {year} {1930})}\BibitemShut {NoStop}%
\bibitem [{\citenamefont {Bethe}(1930)}]{bethe_zur_1930}%
  \BibitemOpen
  \bibfield  {author} {\bibinfo {author} {\bibfnamefont {H.}~\bibnamefont {Bethe}},\ }\bibfield  {title} {\enquote {\bibinfo {title} {Zur {Theorie} des {Durchgangs} schneller {Korpuskularstrahlen} durch {Materie}},}\ }\href {\doibase 10.1002/andp.19303970303} {\bibfield  {journal} {\bibinfo  {journal} {Ann. Phys.}\ }\textbf {\bibinfo {volume} {397}},\ \bibinfo {pages} {325--400} (\bibinfo {year} {1930})}\BibitemShut {NoStop}%
\bibitem [{\citenamefont {Simmermacher}, \citenamefont {Weber},\ and\ \citenamefont {Kirrander}(2023)}]{SimmermacherCh3Kasra}%
  \BibitemOpen
  \bibfield  {author} {\bibinfo {author} {\bibfnamefont {M.}~\bibnamefont {Simmermacher}}, \bibinfo {author} {\bibfnamefont {P.~M.}\ \bibnamefont {Weber}}, \ and\ \bibinfo {author} {\bibfnamefont {A.}~\bibnamefont {Kirrander}},\ }\bibfield  {title} {\enquote {\bibinfo {title} {Theory of time-dependent scattering},}\ }in\ \href {\doibase https://doi.org/10.1039/9781837671564} {\emph {\bibinfo {booktitle} {{Structural Dynamics with X-ray and Electron Scattering}}}},\ \bibinfo {series} {{Theoretical and Computational Chemistry Series}}, Vol.~\bibinfo {volume} {25},\ \bibinfo {editor} {edited by\ \bibinfo {editor} {\bibfnamefont {K.}~\bibnamefont {Amini}}, \bibinfo {editor} {\bibfnamefont {A.}~\bibnamefont {Rouz\'{e}e}}, \ and\ \bibinfo {editor} {\bibfnamefont {M.~J.~J.}\ \bibnamefont {Vrakking}}}\ (\bibinfo  {publisher} {Royal Society of Chemistry},\ \bibinfo {address} {United Kingdom},\ \bibinfo {year} {23 December 2023})\ \bibinfo {edition} {1st}\ ed.,\ Chap.~\bibinfo {chapter} {3}, p.~\bibinfo {pages} {85},\
  \bibinfo {note} {www.rsc.org}\BibitemShut {NoStop}%
\bibitem [{\citenamefont {Prince}(2006)}]{IntTabCryVolC}%
  \BibitemOpen
  \bibinfo {editor} {\bibfnamefont {E.}~\bibnamefont {Prince}},\ ed.,\ \href@noop {} {\emph {\bibinfo {title} {{International Tables for Crystallography Volume C: Mathematical, physical and chemical tables}}}},\ \bibinfo {edition} {2006th}\ ed.,\ \bibinfo {number} {ISBN 978-1-4020-1900-5}\ (\bibinfo  {publisher} {Wiley},\ \bibinfo {year} {2006})\BibitemShut {NoStop}%
\bibitem [{\citenamefont {Salvat}(1991)}]{salvat_elastic_1991}%
  \BibitemOpen
  \bibfield  {author} {\bibinfo {author} {\bibfnamefont {F.}~\bibnamefont {Salvat}},\ }\bibfield  {title} {\enquote {\bibinfo {title} {Elastic scattering of fast electrons and positrons by atoms},}\ }\href {\doibase 10.1103/PhysRevA.43.578} {\bibfield  {journal} {\bibinfo  {journal} {Phys. Rev. A}\ }\textbf {\bibinfo {volume} {43}},\ \bibinfo {pages} {578--581} (\bibinfo {year} {1991})}\BibitemShut {NoStop}%
\bibitem [{\citenamefont {Salvat}, \citenamefont {Jablonski},\ and\ \citenamefont {Powell}(2005)}]{salvat_elsepadirac_2005}%
  \BibitemOpen
  \bibfield  {author} {\bibinfo {author} {\bibfnamefont {F.}~\bibnamefont {Salvat}}, \bibinfo {author} {\bibfnamefont {A.}~\bibnamefont {Jablonski}}, \ and\ \bibinfo {author} {\bibfnamefont {C.~J.}\ \bibnamefont {Powell}},\ }\bibfield  {title} {\enquote {\bibinfo {title} {elsepa—{Dirac} partial-wave calculation of elastic scattering of electrons and positrons by atoms, positive ions and molecules},}\ }\href {\doibase 10.1016/j.cpc.2004.09.006} {\bibfield  {journal} {\bibinfo  {journal} {Comput. Phys. Commun.}\ }\textbf {\bibinfo {volume} {165}},\ \bibinfo {pages} {157--190} (\bibinfo {year} {2005})}\BibitemShut {NoStop}%
\bibitem [{\citenamefont {Debye}(1915)}]{debye_zerstreuung_1915}%
  \BibitemOpen
  \bibfield  {author} {\bibinfo {author} {\bibfnamefont {P.}~\bibnamefont {Debye}},\ }\bibfield  {title} {\enquote {\bibinfo {title} {Zerstreuung von {Röntgenstrahlen}},}\ }\href {\doibase 10.1002/andp.19153510606} {\bibfield  {journal} {\bibinfo  {journal} {Ann. Phys.}\ }\textbf {\bibinfo {volume} {351}},\ \bibinfo {pages} {809--823} (\bibinfo {year} {1915})}\BibitemShut {NoStop}%
\bibitem [{\citenamefont {Werner}\ \emph {et~al.}(2020)\citenamefont {Werner}, \citenamefont {Knowles}, \citenamefont {Manby}, \citenamefont {Black}, \citenamefont {Doll}, \citenamefont {Heßelmann}, \citenamefont {Kats}, \citenamefont {Köhn}, \citenamefont {Korona}, \citenamefont {Kreplin}, \citenamefont {Ma}, \citenamefont {Miller}, \citenamefont {Mitrushchenkov}, \citenamefont {Peterson}, \citenamefont {Polyak}, \citenamefont {Rauhut},\ and\ \citenamefont {Sibaev}}]{Molpro}%
  \BibitemOpen
  \bibfield  {author} {\bibinfo {author} {\bibfnamefont {H.-J.}\ \bibnamefont {Werner}}, \bibinfo {author} {\bibfnamefont {P.~J.}\ \bibnamefont {Knowles}}, \bibinfo {author} {\bibfnamefont {F.~R.}\ \bibnamefont {Manby}}, \bibinfo {author} {\bibfnamefont {J.~A.}\ \bibnamefont {Black}}, \bibinfo {author} {\bibfnamefont {K.}~\bibnamefont {Doll}}, \bibinfo {author} {\bibfnamefont {A.}~\bibnamefont {Heßelmann}}, \bibinfo {author} {\bibfnamefont {D.}~\bibnamefont {Kats}}, \bibinfo {author} {\bibfnamefont {A.}~\bibnamefont {Köhn}}, \bibinfo {author} {\bibfnamefont {T.}~\bibnamefont {Korona}}, \bibinfo {author} {\bibfnamefont {D.~A.}\ \bibnamefont {Kreplin}}, \bibinfo {author} {\bibfnamefont {Q.}~\bibnamefont {Ma}}, \bibinfo {author} {\bibfnamefont {I.}~\bibnamefont {Miller}, \bibfnamefont {Thomas~F.}}, \bibinfo {author} {\bibfnamefont {A.}~\bibnamefont {Mitrushchenkov}}, \bibinfo {author} {\bibfnamefont {K.~A.}\ \bibnamefont {Peterson}}, \bibinfo {author} {\bibfnamefont {I.}~\bibnamefont {Polyak}}, \bibinfo
  {author} {\bibfnamefont {G.}~\bibnamefont {Rauhut}}, \ and\ \bibinfo {author} {\bibfnamefont {M.}~\bibnamefont {Sibaev}},\ }\bibfield  {title} {\enquote {\bibinfo {title} {{The Molpro quantum chemistry package}},}\ }\href {\doibase 10.1063/5.0005081} {\bibfield  {journal} {\bibinfo  {journal} {The Journal of Chemical Physics}\ }\textbf {\bibinfo {volume} {152}},\ \bibinfo {pages} {144107} (\bibinfo {year} {2020})},\ \Eprint {http://arxiv.org/abs/https://pubs.aip.org/aip/jcp/article-pdf/doi/10.1063/5.0005081/16680626/144107\_1\_online.pdf} {https://pubs.aip.org/aip/jcp/article-pdf/doi/10.1063/5.0005081/16680626/144107\_1\_online.pdf} \BibitemShut {NoStop}%
\bibitem [{\citenamefont {Kuhlman}\ \emph {et~al.}(2012)\citenamefont {Kuhlman}, \citenamefont {Sauer}, \citenamefont {S{\o}lling},\ and\ \citenamefont {M{\o}ller}}]{Kuhlman2012a}%
  \BibitemOpen
  \bibfield  {author} {\bibinfo {author} {\bibfnamefont {T.~S.}\ \bibnamefont {Kuhlman}}, \bibinfo {author} {\bibfnamefont {S.~P.~A.}\ \bibnamefont {Sauer}}, \bibinfo {author} {\bibfnamefont {T.~I.}\ \bibnamefont {S{\o}lling}}, \ and\ \bibinfo {author} {\bibfnamefont {K.~B.}\ \bibnamefont {M{\o}ller}},\ }\bibfield  {title} {\enquote {\bibinfo {title} {{Symmetry, vibrational energy redistribution and vibronic coupling: The internal conversion processes of cycloketones}},}\ }\href {\doibase 10.1063/1.4742313} {\bibfield  {journal} {\bibinfo  {journal} {The Journal of Chemical Physics}\ }\textbf {\bibinfo {volume} {137}} (\bibinfo {year} {2012}),\ 10.1063/1.4742313}\BibitemShut {NoStop}%
\bibitem [{\citenamefont {Simmermacher}\ \emph {et~al.}(2019)\citenamefont {Simmermacher}, \citenamefont {Moreno~Carrascosa}, \citenamefont {E.~Henriksen}, \citenamefont {B.~Møller},\ and\ \citenamefont {Kirrander}}]{simmermacher_theory_2019}%
  \BibitemOpen
  \bibfield  {author} {\bibinfo {author} {\bibfnamefont {M.}~\bibnamefont {Simmermacher}}, \bibinfo {author} {\bibfnamefont {A.}~\bibnamefont {Moreno~Carrascosa}}, \bibinfo {author} {\bibfnamefont {N.}~\bibnamefont {E.~Henriksen}}, \bibinfo {author} {\bibfnamefont {K.}~\bibnamefont {B.~Møller}}, \ and\ \bibinfo {author} {\bibfnamefont {A.}~\bibnamefont {Kirrander}},\ }\bibfield  {title} {\enquote {\bibinfo {title} {Theory of ultrafast x-ray scattering by molecules in the gas phase},}\ }\href {\doibase 10.1063/1.5110040} {\bibfield  {journal} {\bibinfo  {journal} {J. Chem. Phys.}\ }\textbf {\bibinfo {volume} {151}},\ \bibinfo {pages} {174302} (\bibinfo {year} {2019})}\BibitemShut {NoStop}%
\bibitem [{\citenamefont {Keefer}\ \emph {et~al.}(2021)\citenamefont {Keefer}, \citenamefont {Aleotti}, \citenamefont {Rouxel}, \citenamefont {Segatta}, \citenamefont {Gu}, \citenamefont {Nenov}, \citenamefont {Garavelli},\ and\ \citenamefont {Mukamel}}]{Keefer2021}%
  \BibitemOpen
  \bibfield  {author} {\bibinfo {author} {\bibfnamefont {D.}~\bibnamefont {Keefer}}, \bibinfo {author} {\bibfnamefont {F.}~\bibnamefont {Aleotti}}, \bibinfo {author} {\bibfnamefont {J.~R.}\ \bibnamefont {Rouxel}}, \bibinfo {author} {\bibfnamefont {F.}~\bibnamefont {Segatta}}, \bibinfo {author} {\bibfnamefont {B.}~\bibnamefont {Gu}}, \bibinfo {author} {\bibfnamefont {A.}~\bibnamefont {Nenov}}, \bibinfo {author} {\bibfnamefont {M.}~\bibnamefont {Garavelli}}, \ and\ \bibinfo {author} {\bibfnamefont {S.}~\bibnamefont {Mukamel}},\ }\bibfield  {title} {\enquote {\bibinfo {title} {Imaging conical intersection dynamics during azobenzene photoisomerization by ultrafast x-ray diffraction},}\ }\href {https://www.pnas.org/content/118/3/e2022037118} {\bibfield  {journal} {\bibinfo  {journal} {Proc.\ Natl.\ Acad.\ Sci.\ U.S.A.}\ }\textbf {\bibinfo {volume} {118}} (\bibinfo {year} {2021})},\ \Eprint {http://arxiv.org/abs/https://www.pnas.org/content/118/3/e2022037118.full.pdf}
  {https://www.pnas.org/content/118/3/e2022037118.full.pdf} \BibitemShut {NoStop}%
\end{thebibliography}%

\end{document}